\documentclass[a4paper,10pt]{article}
\usepackage[T1]{fontenc}
\usepackage[utf8]{inputenc}
\usepackage[italian,english]{babel}
\usepackage[autostyle,italian=guillemets]{csquotes}
\usepackage{multicol}
\usepackage{microtype}
\usepackage{guit}
\usepackage{booktabs}
\usepackage{graphicx}
\usepackage{mathptmx}
\usepackage{amsfonts}
\usepackage{amssymb}
\usepackage{amsmath}
\usepackage{ifpdf}
\usepackage{dcolumn}
\usepackage{booktabs, longtable}
\usepackage{hyperref}
\usepackage{textcomp}
\usepackage{tabularx,array}
\usepackage{ulem}
\usepackage{placeins}
\usepackage{epstopdf}
\usepackage{etex}
\date{}
\title{New Astronomical, Meteorological and Geological Study \\ of Pieve a Nievole (PT)} 
\author{Tasselli, D. \\ TS Corporation Srl - Astronomy and Astrophysics Department\\ Via Rugantino, 71 - 00169 Roma RM - Italy \\ E-mail:diego.tasselli@tscorporation.org \\ \\ Ricci, S. \\ TS Corporation Srl - Meteorogical and Climatic Change Department\\ Via Rugantino, 71 - 00169 Roma RM - Italy \\ E-mail:stefano.ricci@tscorporation.org \\ \\ Bianchi, P. \\ TS Corporation Srl - Geology and Geophysics Department\\ Via Rugantino, 71 - 00169 Roma RM - Italy \\ E-mail:pamela.bianchi@tscorporation.org}
\date{}
\begin{document}
 
 \maketitle
 
 \begin{abstract} 
In this paper we present an analysis of the geological, meteorological and climatic data recorded in Pieve a Nievole (PT) over 24 years and using this data for the establishment of the research structure called "PAN.R.C. - Pieve a Nievole Research Center" and "NGICS - New study of geological and Italian city". These data are compared to check local variations, long term trends and correlation with maen annual temperature. The ultimate goal of this work is to understand long term climatic changes in this geographic area. The analysis is performed using a statistical approach and a particular care is used to minimize any effect due to prejudices in case of lack of data. 
 \end{abstract}

Keyword: atmospheric effects - site testing - earthquake data - geological model - methods: statistical - location: Pieve a Nievole. \\
{\footnotesize This paper was prepared with the \LaTeX \\}
\begin{multicols}%
{2}
\section{\normalsize Introduction} In this paper we present an analysis of astronomical, geological, meteorological and climatic data releaved in Pieve a Nievole (PT), a little town in north Tuscany region. This study is a part of ``NGICS - New study of geological Italian city'' a project by Department of Climatology and Geology of TS Corporation Srl.\\ We present an analysis of measurements obtained from local geological \cite{FBarberi:2013} astronomical and meteorological data  \cite{datimeteo:2013} \cite{eumetsat:2013} and compared in order to check local variations of climatological conditions. 
\subsection{\normalsize Location} Pieve a Nievole is a small town located north of the Tuscan region in Italy. Here are identifiable altimeter data, the geographic coordinates and seismic data.\\ \\
\begin{tabular}{c|c|c}
\hline
\scriptsize \textbf{Latitudine} & \scriptsize \textbf{Longitude} & \scriptsize \textbf{Altitude mt s.l.} \cr
\hline
\scriptsize $43^\circ 52' 21,72$'' N & \scriptsize $10^\circ 48' 9,00$'' E & \scriptsize Min 14 - Max 254  \cr
\hline 
\end{tabular}  \\  \\ \scriptsize {\bf {Geographic Data of Pieve a Nievole}} \\ \\
\normalsize The geological map of the town is shown in Geological and Seismic data section. \cite{mappageologica:2013} \\

\begin{tabular}{lp{0.2\textwidth}}
\hline
\scriptsize \textbf{Seismic Zone} & \scriptsize \textbf{Description}  \cr
\hline 
\scriptsize 3 & \scriptsize Area with average seismic hazard may occur where earthquakes strong enough. The sub area 3 indicates a value of ag < 0,15g. \cr
\hline 
\end{tabular} \\ \\ \scriptsize {\bf {Seismic Data of Pieve a Nievole}}
\section{\normalsize Annual data analysis} \normalsize Summers are generally warm and winters are rigid and often winds blow from the north and may last a long time.\\ The snow appears about one or two times a year, sometimes in large quantities around the appeninical area, moving up to the piedmont areas and rarely up to the coastal areas. The rains dominate the autumn and spring seasons. \\ The following table identifies climate data assigned by Decree of the President of the Republic n. 412 of 26 August 1993.\cite{Tasselli:2011ug}. \\ Characterized by a fairly humid climate especially in inland and coastal areas, the summer can often record the maximum temperature higher than the district, while in winter with the thermal inversion and the particular conditions in which it occurs, it manages to be the coldest city in the area (obviously compared to the cities in which data is monitored). \\ \\
\begin{tabular}{c|c}
\hline
\scriptsize \textbf{Climatic Zone} & \scriptsize \textbf{Day Degrees} \cr
\hline 
\scriptsize D & \scriptsize 1.708 \cr
\hline 
\end{tabular} \\ \\ \scriptsize {\bf {Climatic Parameter of Pieve a Nievole}} \\
\section{\normalsize Meteo-Climatic Parameter} 
\normalsize In this section we describe air temperatures (T), Dew Point, Humidity, Pressure, Day Time and Night Time Variation, Rain's Days and Fog's Day, obtained by an accurate analysis of the meteorological data from local data by archive \cite{datimeteo:2013}. We analyzed the parameters given in Table 1 and 2. \\Should be noted that the values considered are related to the last twenty-four-year average and made available for the period 1990-2014. \\ \\
\begin{tabular}{lp{0.10\textwidth}} 
\hline
{\footnotesize Average Annual Temperature}&  {\footnotesize $14,98 ^\circ C$}  \cr
{\footnotesize T average warmest (ago-03)}& {\footnotesize $27,75 ^\circ C$} \cr
{\footnotesize T average coldest (feb-12)}&  {\footnotesize $1,55 ^\circ C$} \cr 
{\footnotesize Annual temperature range}&  {\footnotesize $10,10 ^\circ C$}  \cr
{\footnotesize Months with average T > $20 ^\circ C$} & {\footnotesize  89} \cr
{\footnotesize Total rainfall 1990-2014 [mm]}& {\footnotesize 26252,48} \cr
{\footnotesize Rain Days }&  {\footnotesize 1868}  \cr
{\footnotesize Fog Days }&  {\footnotesize 1052}  \cr
{\footnotesize Storm Days} &  {\footnotesize 98}  \cr
{\footnotesize Rain/Storm Days }& {\footnotesize 899}  \cr
{\footnotesize Rain/Snow Days}&  {\footnotesize 19}  \cr
{\footnotesize Rain/Fog Days} & {\footnotesize 260}  \cr
{\footnotesize Rain/Thunder/Fog Days} & {\footnotesize 81}  \cr
{\footnotesize Snow Days} & {\footnotesize 9}  \cr
{\footnotesize Wind Speed max Km/h (Mar-1995)} & {\footnotesize 33,06}  \cr
{\footnotesize Wind Speed min Km/h (Nov-2007)} & {\footnotesize 7,00}  \cr
{\footnotesize Rain max mm (Oct-1990)} & {\footnotesize 637,00}  \cr
{\footnotesize Rain min mm (Jul-2009)} & {\footnotesize 1,00}  \cr
{\footnotesize Earthquake Min (2003/06/01)} & {\footnotesize 1,0 Mw}  \cr
{\footnotesize Earthquake Max (2014/10/06)} & {\footnotesize 2,4 Mw}  \cr
{\footnotesize Earthquake Deep Min (2004/10/06)} & {\footnotesize 3,0 Km}  \cr
{\footnotesize Earthquake Deep Max (2005/05/13)} & {\footnotesize 52,0 Km}  \cr
\hline
\end{tabular} 
\\ \\ \mbox{\bf{\footnotesize  Parameter of this Study}}
\subsection{\normalsize Solar Radiation Territory}  \normalsize In the study area the values of solar radioation are increasing.\\ In fact the data irradiation of territory taken from the parameters and the data prepared by the European Union, demonstrate the trend of irradiation for the municipality, visible in next table: \\
\begin{tabular}{|c|c|c|c|}
\hline
     \bf {\footnotesize Month} & \bf {\footnotesize DNI}& \bf {\footnotesize Month} & \bf {\footnotesize DNI}\cr \hline
       {\footnotesize Jan} &  {\footnotesize 12,05} & {\footnotesize Feb} &  {\footnotesize 14,32} \cr
       {\footnotesize Mar} &   {\footnotesize 19,44} & {\footnotesize Apr} &  {\footnotesize 21,54} \cr
       {\footnotesize May} & {\footnotesize 26,24} & {\footnotesize Jun} &  {\footnotesize 28,69} \cr
       {\footnotesize Jul} &   {\footnotesize 29,11} &  {\footnotesize Aug} & {\footnotesize 24,99} \cr
       {\footnotesize Sep} & {\footnotesize 20,69} &  {\footnotesize Oct} & {\footnotesize 15,27} \cr
      {\footnotesize Nov} & {\footnotesize 11,67} & {\footnotesize Dec} &  {\footnotesize 9,67} \cr \hline
     \bf {\footnotesize Year} & \bf {\footnotesize 233,68} & & \cr
\hline
\end{tabular}
\\ \\ \mbox{\bf{\footnotesize Direct Normal Irradiance (Wh/m$^2$/day) }} \\ \\
\normalsize The weather data and the graphs show extrapolated for the territory covered by the study, including a radiation in the range between 1100 and 1200 kWh /1kWp as map prepared by the European Union \cite{UE:2013} and visible in figure 17, characterized in over the months, irradiation presented in the graph in Figure 18 and 19, which shows the data of the table above, which shows the territory of Pieve a Nievole, a total irradiance Annual of 233,68 Wh/m$^2$/day.
\subsection{\normalsize Temperature} \normalsize In this section we describe air temperatures (T) obtained by an accurate analysis of the meteorological data. The year average temperature has a tendency to remain stable throughout the study period; next graphics show this evidence: \\ \\
\includegraphics[width=0.49\textwidth{}]{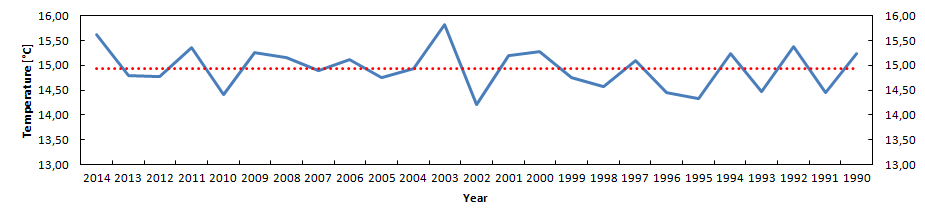} 
{\scriptsize \bf \textit{\textit{{\footnotesize Relationschip of Year Temperature and Average 1990-2014}}}}. \\
 \\The values are calculated considering the entire measurement period (1990-2014), drawing on data from the Annuals published by "Il Meteo.it" \cite{datimeteo:2013} for the period 1990-2014. \\ The study shows sudden periods of descent and ascent in the shape of a step from 1990 to 1995, going dows at the end of 1994 and going up around the average in 1997. In the period 1997-2003 the average monthly temperatures were around the average reaching the maximun peak in the August 2003, and the lowest peak in January 2002. \\Since 2003 there has been a gradual rise, with the average temperature recorded was lower than the average, to then rise constantly until 2014, the year of the end rewriting of the present study.\\ \\ \includegraphics[width=0.49\textwidth{}]{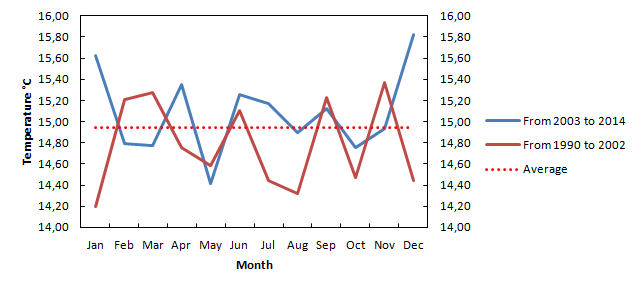} 
 {\scriptsize \bf \textit{\textit{{\footnotesize Variation of Month Temperature about year 1990-2002 and 2003-2014}}}} \\ \\ The variation in average monthly temperaturees, visible in the previous grph divided into the periods 1990-2002 and 2003-2014, shows the trend of average temperatures. A detailed analysis highlights the following: 
 \begin{itemize}
 	\item From 1990 to 2002: the variation of $1.00^\circ C$ occured in the period Apr-May;
 	\item From 2003 to 2014: two period were highlighted where changes were recorded in the Sep-Oct and Nov-Dec. 
 \end{itemize}
 The graphs shown in Figures 3 and 4 show the performance of the maximum and minimum temperatures proportional to the average of the period of study (1990-2014). The study shows the following trends:
\begin{itemize}
\item Minimum temperatures: there was an increase of the values recorded with a higher peak in February 2012, with $1,55^\circ C$;
\item Maximum temperatures: evidence for the period 2007-2009 is an increasing trend compared to the 1990-2014 average, while evidence a decrease in the period 1990-1997. The maximum temperature was $27.75^\circ C$ in August 2003;
\end{itemize} 
Table 5 shows the trend of the points of maximum temperature expected in the summer quarter (June-July-August) and are shown in Figure 14. Study shows that the month of August 2003 recorded the highest average maximum value in a context in which the entire month recorded in the period 1990-2014 values always above average. Alway Table 6 shows the trend of the points of minimum temperature expected in the winter quarter (January-February-March) and are shown in Figure 15. Study shows that the month of February recorded the highest average minimum value, in a context in which the entire month recorded in the period 1990-2014 values always above average.
Studying in particular the months with the values of minimum temperature and maximum minors, the following is noted:
\begin{itemize}
\item the month of August (characterized by the presence of the highest value observed in maximum temperatures), shows that trends in temperature has been getting consistently below average for the period 1990 to 2014, while there were two exceedances of this value over the years 2003 and 2012, in agreement with what evident from the graph in Figure 12;
\item the month of February (characterization from the the lowest minimum temperature for the period of study), shows that the trend of temperatures has always been, in agreement with what reported in the graph in Figure 13.
\end{itemize}
\subsection{\normalsize Blackbird days} The day from 29 to 31 January called in the Italian tradition "Blackbird days" are the coldest days of the year. The study showed an upward trend in the reference period, showing an average of $2.22^\circ C$. \\The following grap shows this trend.\\
\includegraphics[width=0.49\textwidth{}]{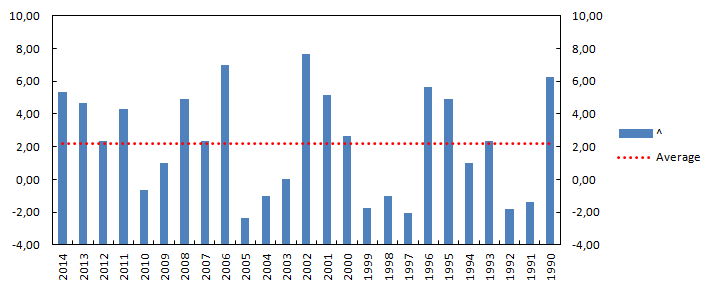} {\scriptsize \bf \textit{\textit{{\footnotesize Blackbird Days from 1990 to 2014}}}}.
\subsection{\normalsize Dew point} In this section we describe Dew point obtained by an accurate analysis of the meteorological data. This study   showed an average on the dew point of $10.07^\circ C$ showing that for 14 years the recorded values were above average, as shown in Figure 10, while at the monthly level they were recorded in the study period 6 months over the average, as show in the following graph: \\ \\
\includegraphics[width=0.49\textwidth{}]{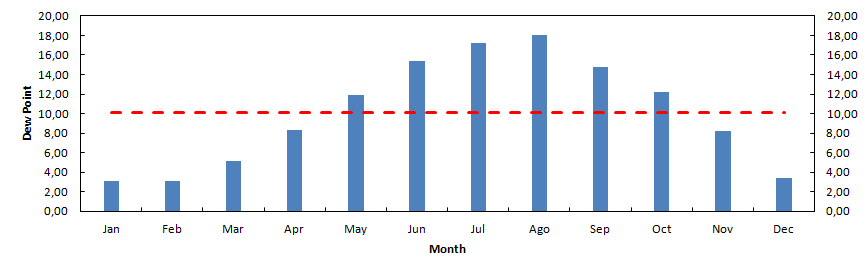} {\scriptsize \bf \textit{\textit{{\footnotesize Dew Point for months from 1990 to 2014}}}}. 
\subsection{\normalsize Humidity} In this section we describe humidity obtained by an accurate analysis of the meteorological data. The study highlights a gap in the annual humidity values equal to 74,17\%, calculated according to this formula:
\begin{equation}\frac{av}{am}\end{equation} 
{\footnotesize \textit{ {\bf av} = annual value, {\bf am} = average moisture 1990-2014}} \\ \\ The graph in Figure 8 shows the trend at the annual level while in the following graph it shows it at a monthly level. \\ \\
\includegraphics[width=0.49\textwidth{}]{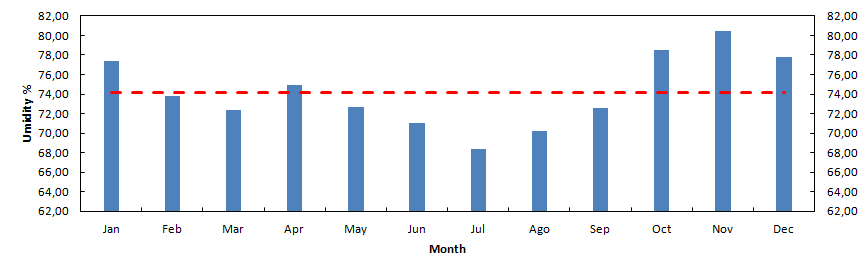} {\scriptsize \bf \textit{\textit{{\footnotesize Humidity Months from 1990 to 2014}}}}. \\ \\
The study showed that in the annual period there were 14 years out of a total of 25, while at the monthly level the months above average were 5.
\subsection{\normalsize Pressure} \normalsize In this section we describe pressure obtained by an accurate analysis of the Pressure data. \\ The analysis of the data showed an average for the study period of 1015 hPc, a strong monthly increase in the values recorded in Jannuary, Febrary and December, and a monthly decrease in the months of April, May, June and July as shown in the following chart: \\ \\
\includegraphics[width=0.49\textwidth{}]{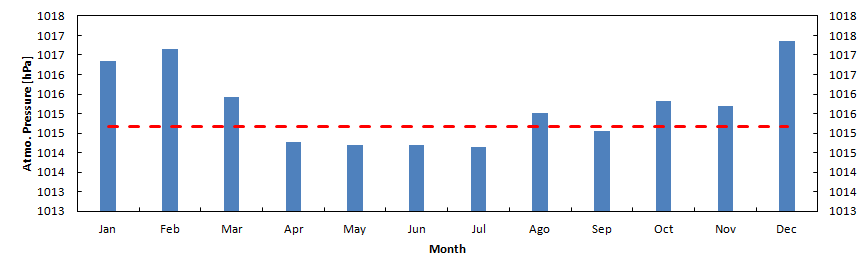} {\scriptsize \bf \textit{\textit{{\footnotesize Pressure Months from 1990 to 2014}}}}. \\ \\
Analyzing the data at an annual level, we highlight the following:
\begin{itemize}
\item an increase in atmospheric pressure over the years: 1990, 1991,1992 (year with the highest level of pressure throughout the study period), 1994, 1995, 1997, 1998, 1999, 2000, 2003, 2004, 2006, 2007, 2008 and 2011;
\item a decrease in atmospheric pressure over the years: 1996, 2001, 2002 (the year with the lowest level of pressure throughout the study period), 2009, 2010 2013 and 2014.
\end{itemize}
In Figure 9 we can see the Pressure Graphics for this study. 
\subsection{\normalsize Day time and night time variation} In this section we describe number of Day time and night time variations obtained by an accurate analysis of the meteorological data. 
The annual averages of the differences between day time and night time temperatures $\Delta T$ have been computed and the results are reported in Table 3. Also in next figure, we can see the plot of oscillations of the $\Delta T$ seem to reduce the amplitude during the years. \\ The difference of temperatures, with an average difference of $10,10^\circ C$ (see table 2) and next figure show these effects. \\ \\ \includegraphics[width=0.49\textwidth{}]{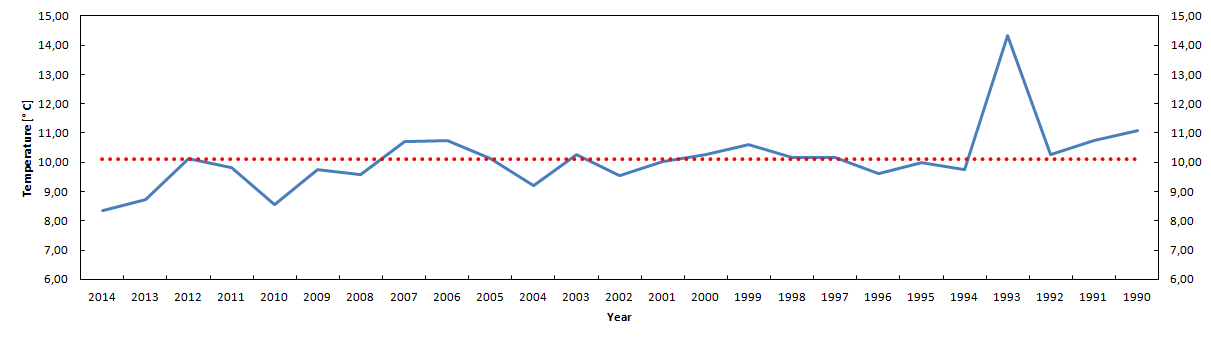}
{\scriptsize \bf \textit{\textit{{\footnotesize Day time and night time variations (solid line) and Average Day time and Night time variations 1990-2014 (dotted line)}}}}
\subsection{\normalsize Rain's Days} In this section we describe number of Rain's days obtained by an accurate analysis of the meteorological data. \\  The study showed that if on the one hand at the monthly level the rain were lower than the average - value calculated in 232,17 mm - for a total of 7 months as shown in the following graph, \\ \\ \includegraphics[width=0.49\textwidth{}]{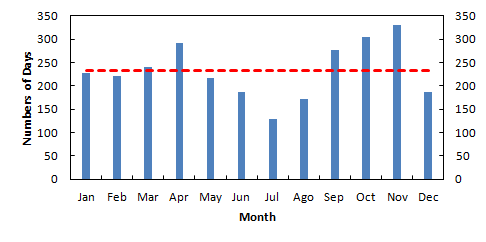} 
{\scriptsize \bf \textit{\textit{{\footnotesize Days Rain for Month from 1990 to 2014}}}}. \\ \\ on the other there was an increase in the phenomena extremes at monthly level, as shown in figure 16 where the amounth of ainfall in mm dropped for each  single month in the reference period 1990-2014 can be viewed.
The increase in intense rainfall of short duration boosts the risk of flash-food, fast flooding of an area circumscribed morphologically, due to the fast ``saturation'' of the surface soil that is no longer able to absorb the rain. The episodes of intense precipitation can also determine the phenomena of surface runoff of rainwater with a possible increase in flooding, but also the risk of water pollution (pollutants from agricultural and road runoff). \\ The study period showed an abnormal increase in rainfall. The pie chart shows the ratio between the total mm of rain measured in a year and the total rainfall measured in the period 1990-2014, according to this formula: \begin{equation}Total Rain=\frac{a}{b}x100\end{equation} 
{\footnotesize \textit{{\bf a} = annual rainfall value, {\bf b} = total value of the rain period}} \\ \includegraphics[width=0.49\textwidth{}]{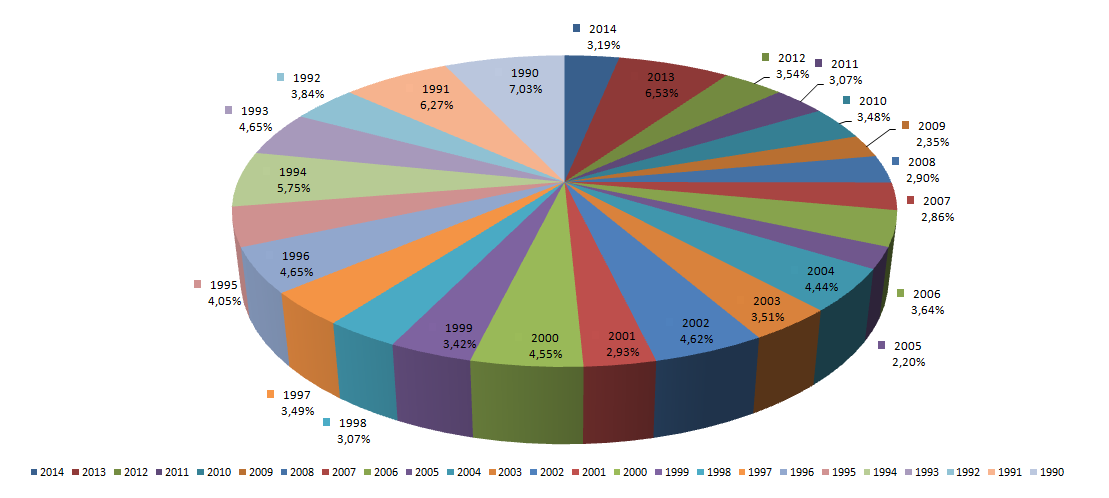}
{\scriptsize \bf \textit{\% of Rain for this study}}.
\\ \\ In Figure 6 and 7 we can see the total of rain/year in this study. The values are visible in the table 3. Next pie chart evidence type of rain for this study. \\ \\
\includegraphics[width=0.5\textwidth{}]{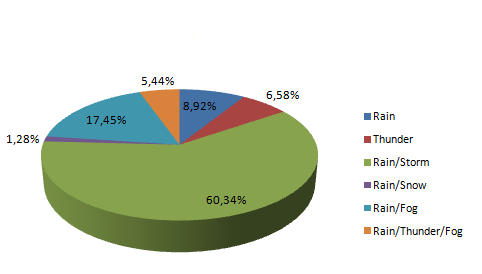}
{\scriptsize \bf \textit{Type of Rain for this study}}. \\ \\ Figure 16 evidence the total rain in mm, for only month. \\ The study showed an average annual rainfall of 1050,10 mm, with an increase in the years: 1990 (the highest year period mm of rain fell), 1991, 1993, 1994, 1996, 2000, 2002, 2004, and 2013; while at the monthly level the average of 232,17 mm was exceeded in the months of April, September, October and November as shown in the graph at the beginning of this paragraph.
\subsection{\normalsize Fog's days} In this section we describe number of Fog's days obtained by an accurate analysis of the meteorological data. \\ The study showed a trend increase in the presence of days with fog and evidence that October is the first month for number of Fog's day and July are the laster. Figure 7 we can see the number of Fog days by this study. \\ \\
\includegraphics[width=0.49\textwidth{}]{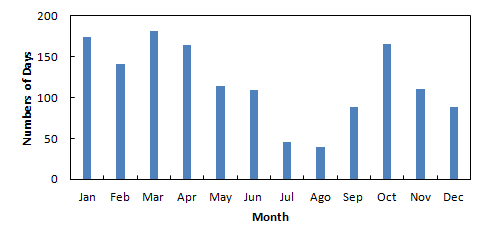}
{\scriptsize \bf \textit{Days Fog for Month from 1990 to 2014}} 
\subsection{\normalsize Wind speed} 
\normalsize In this section we describe the wind speed. \\ In the above image you can see the map of wind speed insistent on the territory of Pieve a Nievole, as seen from the map generated from Atlas wind \cite{atlantevento:2013}. \\ The study of daily wind speed has allowed to estimate on a monthly basis throughout the period included in this study:
\begin{itemize}
	\item a decrease below the average of the period 1990-2014 in the minimun values of wind speed, with values ranging between 16 Km/h (in January), and 19 Km/h in February;
	\item an increase above the average maximum wind speed, betweenthe minimuof 22 km/h in May and 29 km/h in December.
\end{itemize} 
\normalsize The study of the period showed that the month of January recorded the greatest variation between the minimum and maximum wind speed, as clearly shownn in the graph of Figure 11. \\ \\ 
\includegraphics[width=0.49\textwidth{}]{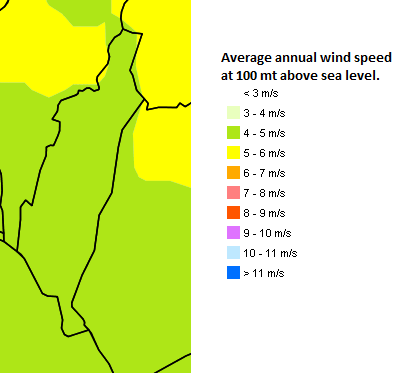} \\ 
{\scriptsize \bf \textit{Wind Speed Atlant of Pieve a Nievole - PT \cite{atlantevento:2013}}}  

\section{\normalsize Geological and Earthquake data} In this section we describe the Geological and Earthquake data. \\
\includegraphics[width=0.49\textwidth{}]{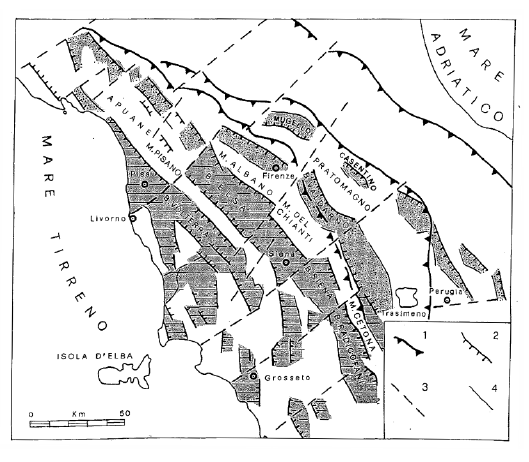} \\
{\scriptsize \bf \textit{Fig.1: Distribution of the main neogenic and quaternary basins of the Northern Apennines. 1=main overlapping fronts; 2=main faults at the edges of the basins; 3=transverse tectonic lines; 4=minor faults at the edge of the basins (BOSSIO et all. 1992).}} \cite{mappageologica:2013} \\ \\ 
The geological area of the present study belongs to the central zone of the orogenic chain of the Apennines northern and is an integral part of the perimediterranean deformation belt, developed in neogenic times and consisting of a complex structure of groundwater and thrust formed in relation to several tectonic phases.\\ The Apennines is a pitched chain characterized by the superposition of paleogeographic elements more interiors on more external elements; the tectonic story that led to his training has developed continuously and is still in progress, as demonstrated by the continuous seismicity.\\ In its evolution it is possible to distinguish some periods (tectonic phases) in which the intensity of the deformations is particularly high and such as to leave a trace so summarizable stratigraphic:
\begin{itemize}
\item From the Cretaceous to the Middle Eocene (oceanic phase) the progressive closure of the paleo occurs Ligurian-Piedmontese Ocean. The main tectonic phase refers to the middle Eocene and is testified by the important one discrepancy between the Epiligure Succession and the previously deformed Liguridi underlying;
\item The deformative phases that followed Oligocene on, occurred in a framework very different geodynamic (continental phase) represented by a collisional regime e post-collisional in which they were progressively involved in the deformation le sequences of the Tuscan and Umbrian foreland with the overhanging deposits of the foredeep.
\item During the Oligo-Miocene collision the Ligurian Units override the Units Tuscan and Umbrian Marchigiane. The migration of the deformation front towards the outside has been accompanied by the translation of the Ligurian Units towards the northeast, generating the presence of holistostromes with prevailing Ligurian lands interspersed in flyschoid deposits. The displacement of the deformation front during the Miocene and the Pliocene subsequently involved more domains exteriors of the Apennine foreland (Umbrian domain - Marche and Padano) up to determine the current configuration of the north - Apennine chain;
\item From the upper Tortonian in the inner part of the Apennine chain developed gods basins (eg. Mugello basin, Florence-Prato-Pistoia basin, Valdarno basin), which interpreted as generated in relaxing regime. This scheme was explained by an evolutionary model of the chain that included the migration of the eastward compressive front with the consequent establishment of an extension regime in the more areas internal. Recent studies hypothesize reactivations in compression of thrusts, even crustals, during the late Miocene, Pliocene and Pleistocene (Boccaletti and Sani, 1998).
\end{itemize}
\includegraphics[width=0.49\textwidth{}]{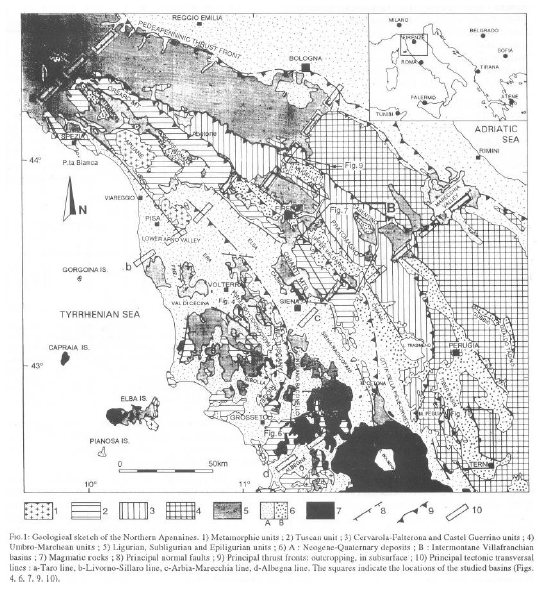} \\
{\scriptsize \bf \textit{Fig. 2: Geological framework of the Northern Apennines (from Boccaletti et al. 1996)}} \cite{mappageologica:2013} \\ \\ The main tectonic element of the Florence-Prato-Pistoia basin is represented by the system of faults along the line Prato - Fiesole, which consists of a sub-parallel beam of normal faults, arranged in steps and oriented around NW-SE with SW immersion. This system delimits the eastern margin of the basin and is formed by several sub-parallel faults arranged in tiers that in the study area they are completely buried below the lake and alluvial sediments.\\ The Florence-Prato-Pistoia basin has a fluvial-lacustrine nature attributable to the Pleistocene lower (the upper Pliocene is not excluded), following the formation of a semi-structure-graben regulated by the Fiesole fault system. The maximum depth of the basin is 600 m in the Prato area and gradually decreases towards E to reach the minimum depthin the area of Florence city, where about 50 m of sediments are found.\\ The current geological configuration of the territory in question represents the final phase of a complex process of transformation of vast geological environments that have over time suffered a slow and deep transformation. The area considered belongs to the geological and structural area of the northern Apennines, reflecting the salient features; there are two styles superimposed tectonics: a plicative-compressive one, related to the Apennine orogenesis, and a rigidistensive one related to the post-orogenic phase.\\
The first of the two deformation phases, which produced the Apennine pitch structure Northern, caused the wrinkling of the Tuscan Series and the overthrusting of the Liguridi;\\
the next originated a structure consisting of a series of raised blocks ("horst") and of depressed areas ("graben") arranged parallel to each other and delimited by systems of faults directed with Apennine orientation (NW-SE).\\	
At the end of the Pliocene (Astiano), about 2 million years ago, a rapid regression leads to definitive establishment of a lake environment that includes the depressions of Fucecchio and Bientina communicating with each other: a large lake extends from the Apennine foothills to the ridge of the Montalbano lapping the Pisan Mountains, bordered to the south by the Pliocene hills of Cerreto Do you drive. This basin is affected by a deposition of fluoro-lacustrine villafranchian sediments a mostly fine grain size.
\subsection{\normalsize Geomorphology} 
The area under examination was analyzed from a geomorphological point of view by photointerpretation in stereoscopic vision and with verification on the ground of the acquired data, in order to discriminate e recognize the set of forms and phenomena that may be of practical interest in relation to them of the realization of the works in project. The elements derived from the photointerpretation result in good agreement with what was found directly on site, in any case the campaign checks and the the results of the geognostic investigations made it possible to complete the indications provided by the photointerpretation, defining a detailed geomorphological picture of the area, represented within the geomorphological cartography attached to the present project.	
\subsection{\normalsize Description of the Formsand Processes}
Geomorphological analysis identifies and recognizes the various physical forms produced by the agents morphogenetics such as gravity, surface water flow, chemical dissolution the action of wind, of the sea and the work of man. This type of survey allows to reconstruct the dynamic picture of land changes that occur slowly or quickly a depending on the prevalence of physical dynamics over those induced by human activities.\\	The legend used for cataloging and the description of the geomorphological phenomena was then constructed by differentiating the forms due to the various morphogenetic agents in the section examined:
\begin{itemize}
\item Forms and processes due to gravity;
\item Forms and processes due to water flow;
\item Anthropic forms and processes.
\end{itemize}
Within these main categories, the activity of the detected forms is taken into account have been distinguished, when possible and significant, in active, quiescent and inactive forms; the first indicate phenomena that may constitute conditions of real risk such as to impose interventions of made safe, even if in a different degree, while inactive forms can constitute situations of potential danger that can eventually degenerate during events exceptional meteorological or improper soil transformation operations. Gravitational processes include landslide and soliflux phenomena, mapped forms are edges and landslide accumulations (quiescent), edges of degradation escarpments and morphological slopes with edge rounded.\\
The areas have been mapped as regards the forms linked to the flow of water with watercourses in depth, ditches in erosion, edges of river and torrent escarpment, hydrographic networks.\\ The anthropic processes and forms have been divided into: embankment and / or carry slope edges embankments (roads and railways) and embankments and artificial barriers artificial waterways.
\subsection{\normalsize Side shapes due to gravity}
\subsubsection{\normalsize Landslide accumulation}
These are accumulations of material generated by movements on the slope. In the investigated area were detected quiescent landslides with rototranslative movement. Quiescent forms are those	with gravitational processes not in place, recent, and probably not yet completely stabilized.	
\subsubsection{\normalsize Surface affected by soliflux or creep}
Soliflux and creep are types of surface movement that can be correlated to plastic deformations of the soil, which can also occur on limited slopes. The slowness of this type of process and the lack of obvious cutting surfaces or detachment niches shapes the terrain into an inconspicuous manner and therefore also the area delimitation of the phenomenon sometimes becomes difficulty. For this reason the phenomenon is not usual, but it is highlighted through the use of a discrete symbology. Areas subject to flow should be treated with caution a due to the possible evolution of the phenomena, which can be equated with active landslides of blanket. In the section in question this type of instability is not directly interfering with the route motorway project.
\subsubsection{\normalsize Degradation slope}
These are rough slopes that can be determined by various factors, including simple ones lithological variations or particular structural features of rock masses. This form is generally indicative of precarious equilibrium conditions, or of situations that could give place to instability in the event of incautious human interventions, seismic actions or changes in the erosive capacity of surface waters.
\subsubsection{\normalsize Morphological slope with rounded edge}
These are modest variations of slope of a slope that determine the formation of a small step, generally with a very well rounded edge and therefore testify to one current situation of substantial geomorphological equilibrium; such forms may be useful for focus on the recent evolution of the landscape. River and slope forms due to water surface.\\ The evolutionary geology of the study are is highlighted below, as also indicated in the paper presenting the new research infrastructure as shown below. \cite{Tasselli:2018} \\
From a geological point of view, you can not look at the area of the Pieve a Nievole town, without inserting it into a wider contest that includes the whole surrounding flat area. \\ The area affected by the project is a flat area characterized by the presence of the large {\itshape "palude of Fucecchio"} whose drainage occurred in various historical epoche. This type of area is frequent in Tuscany, and is a phenomenon resulting from the genetic phenomenon due to the reduction of the comprehensive efforts that have originated in the Apennine chain. The beginning of the evolution of the area can be summarized as follows: \\
\begin{itemize}
\item \textbf{Middle-upper Pliocene (from 4 to 2 Million Years ago)}\\
If Italy is on a larger scale conceived as already defined in the main orographic characteristics, the situation seen in the area covered by the present study is different, as this area consists of a slope with a slight slope crossed by rivers that descend from the north.\\
\end{itemize}
\begin{itemize}
\item \textbf{Lower Pleistocene (Villafranchiano Superiore from 2 to 0.7 million years ago)}\\ This period is characterized by the progressive lifting of the area due to a progressive subsidence, linked to the phenomena of migration to the east of the comprehensive efforts responsible for the birth of the Apennines. The consequence of these events was the birth of a lake in which the paleo rivers brought sediments, while the ground due to the effect of subsidenze continued to lower until it touches 400 meters below sea level. \\ At the end of this period, probably due to a glacial event, the sea level was lowered, until it came to an offset of the coast, not unlike the current configuration.\\ In this period also the lakes present disappeared, and specifically in the area affected by the project, this phenomenon due to the lower subsidence compared to the sedimentary contribution, determined a reduction rather than a disappearance of the lake basin of Pescia-Empoli In this period, the residual \textit{"Fucecchio lake area"} turned into a swamp.
\end{itemize}	
\begin{itemize}
	\item \textbf{ Middle-upper Pleistocene (from 700,000 to 8,800 years ago)}\\
	In this period the area still affected by elevation, assumed the current configuration. \\ There were no substantial orographic changes except for a drift of the swamp that occupied the eastern side.	
\end{itemize}
\begin{itemize}
	\item \textbf{ Holocene (from 8,800 years to current date)}\\
	In this perimeter the variations are almost all of an anthropic nature. In fact, during the \textit{"Leopoldino period"} we can see the reclamation of the Fucecchio marsh which has reduced its extinction and regulated the waters.
\end{itemize}
\subsection{\normalsize Earthquake}
\normalsize The municiplay of Pieve a Nievole is characterized by a territory with low seismicity, \\ \\ \includegraphics[width=0.49\textwidth{}]{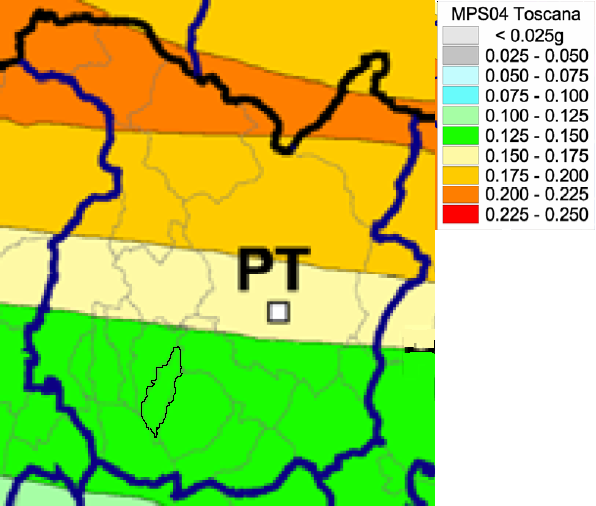} \\
{\scriptsize \bf \textit{Seismic Hazard Map (MPS) of Tuscany. The acceleration values refer to a return of 475 years (INGV2004). The city of Pieve a Nievole is highlighted in black on the map.}} \cite{INGV:2004} \\ \\
in fact in table A attached to the Decree of 01/14/2008 draw up by the Ministry of Infrastructures, the seismic hazard  estimates are highlighted and from these the response spectrum was determined elastic (horizontal and vertical) of the seismic actions highlighted in the following graph.
\includegraphics[width=0.49\textwidth{}]{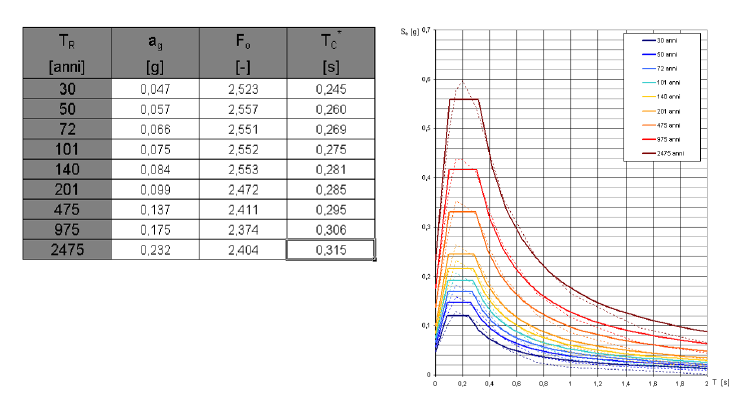} \\
{\scriptsize \bf \textit{Chart show the Ag, Fc, Tc values for the return period TR in years. Solid lines are the normative spectra, dashed line the spectra of S1-INGV project from which they are derived.}} \cite{INGV:2004:ref2} \\ \\
In next figure we can see the number of Earthquake in Pieve a Nievole, and table evidence the number of Earthquake event by year. \\ \\
\includegraphics[width=0.5\textwidth{}]{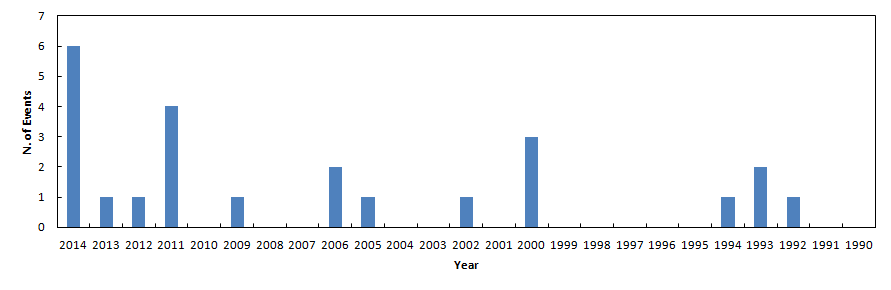} \\
{\scriptsize \bf \textit{Number of Earthquake for 1990 to 2014}} \\
\begin{tabular}{|r|c|c|c|}
\hline
\scriptsize \textbf{Year} &\scriptsize \textbf{N. of Events} &\scriptsize \textbf{ Year} & \scriptsize \textbf{N. of Events}\\
\hline
\scriptsize 2014 & \scriptsize 6 & \scriptsize 2013 & \scriptsize 2 \cr
\scriptsize 2012 & \scriptsize 1 & \scriptsize 2011 & \scriptsize 4 \cr
\scriptsize 2010 & \scriptsize 0 & \scriptsize 2009 & \scriptsize 1 \cr
\scriptsize 2008 & \scriptsize 0 & \scriptsize 2007 &\scriptsize 0 \cr
\scriptsize 2006 & \scriptsize 2 &\scriptsize 2005 & \scriptsize 1 \cr
\scriptsize 2004 & \scriptsize 0 & \scriptsize 2003 & \scriptsize 0 \cr
\scriptsize 2002 & \scriptsize 1 & \scriptsize 2001 & \scriptsize 0 \cr
\scriptsize 2000 &\scriptsize 3 & \scriptsize 1999 & \scriptsize 0 \cr
\scriptsize 1998 & \scriptsize 0 & \scriptsize 1997 & \scriptsize 0 \cr
\scriptsize 1996 & \scriptsize 0 &\scriptsize 1995 & \scriptsize 0 \cr
\scriptsize 1994 & \scriptsize 1 & \scriptsize 1993 & \scriptsize 2 \cr
\scriptsize 1992 & \scriptsize 1 &\scriptsize 1991 & \scriptsize 0 \cr 
\scriptsize 1990 &\scriptsize 0 & & \\
\hline
\end{tabular} \\ \\
\scriptsize {\bf {Number of Earthquake in Pieve a Nievole by Year}} \\ \\
\normalsize The study evidence that the major number of Earthquake are production on deep from 5 to 10 Km. Next graphics evidence this. \\ \\
\includegraphics[width=0.5\textwidth{}]{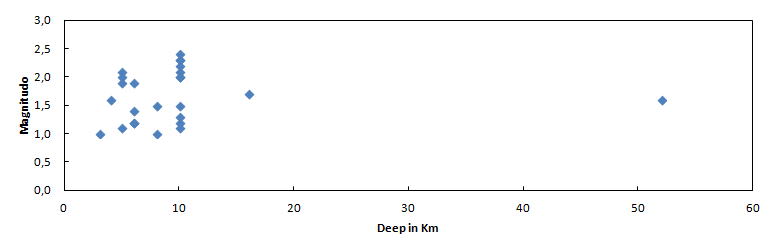} \\
{\scriptsize \bf \textit{Magnitudo/Deep for 1990 to 2014}}
\section{\normalsize Conclusion} 
We presented for the first time an analysis of geological and longterm temperature data directly obtained from Pieve a Nievole local meteorological site, inside urban concentration and well above the inversion layer. \\
The data analysis above all those related to annual rainfall, to their regimen and the change of these indices over time indicates that the area is recorded increase the contribution of rainfall and its concentration in the most rainy periods. \\
The extremes of this trend over time if confirmed, would lead to a radical change in climate of the area that would no longer be characterized by the absence of the dry season, but rather by a rainy season with heavy and often heavy rainfall and a dry season with characteristics similar to those of the dry seasons of the Mediterranean climate. \\ The inversion of Pieve a Nievole arouses contrasting effects among citizens, as it is a phenomenon particularly popular in the middle of a hot summer when at least in the early morning you are able to get much lower temperatures and refreshing, while in winter it is seen as a phenomenon inconvenient dates the low temperatures and possible fog and frost that usually occur in parallel. \\
Given the geology structure of the rocks and the territory, and highlighted the amount of railfallin the period presented in this study (highlighted in section 3.7 and in Table 3), it is recommended further attention in the monitorng of climatic phenomena, in order to prepare the necessary and appropriate measure to protect the population and existing infrastructure.
\section{\normalsize Acknowledgments}
We would like to thank Dr.ssa Filomena Barra for her helpful suggestions and support to made the paper more complete. The constructive comments are highly appreciated. 
\end{multicols}
\newpage
\begin{figure}
\begin{center}
\includegraphics[width=0.8\textwidth]{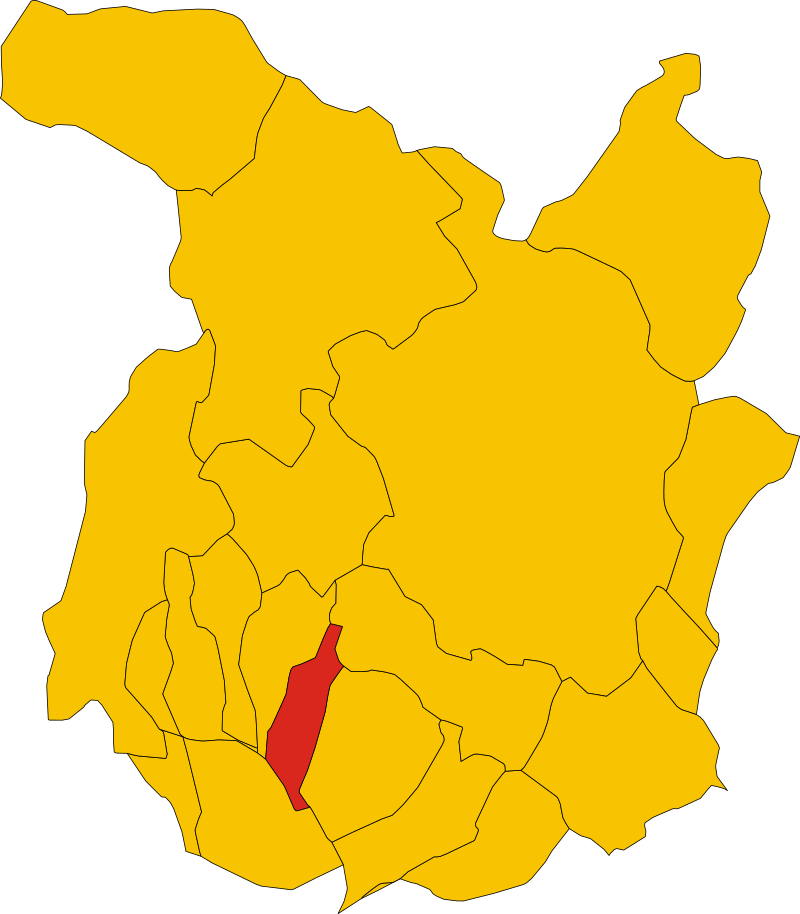}
\caption{Identification of Pieve a Nievole in the Pistoia's  province}
\end{center}
\end{figure}
\clearpage
\begin{table}
\normalsize
\caption{Comparison of temperature on decadal scale from Jannary to June} 
\normalsize
\begin{tabular}{|c|c|c|c|c|c|c|c|c|c|c|c|c|}
	\hline
	\multicolumn{ 1}{|c}{Year} & \multicolumn{ 2}{c}{Jan} & \multicolumn{ 2}{c}{Feb} & \multicolumn{ 2}{c}{Mar} & \multicolumn{ 2}{c}{Apr} & \multicolumn{ 2}{c}{May} & \multicolumn{ 2}{c|}{Jun} \\
	
	\multicolumn{ 1}{|c|}{} &        Min &        Max &        Min &        Max &        Min &        Max &        Min &        Max &        Min &        Max &        Min &        Max \\
	\hline
	2014 &         -1 &         15 &          2 &         17 &          3 &         21 &          3 &         22 &          7 &         26 &         10 &         34 \cr
	
	2013 &         -2 &         15 &         -3 &         15 &          0 &         16 &          6 &         25 &          7 &         24 &         11 &         32 \cr
	
	2012 &         -4 &         14 &         -7 &         17 &          1 &         24 &          4 &         27 &          5 &         26 &         10 &         33 \cr
	
	2011 &         -3 &         15 &         -1 &         15 &         -2 &         18 &          4 &         23 &          7 &         30 &         14 &         30 \cr
	
	2010 &         -3 &         13 &         -4 &         16 &          0 &         18 &          3 &         24 &          9 &         26 &         13 &         32 \cr
	
	2009 &         -4 &         15 &         -3 &         16 &         -2 &         18 &          7 &         22 &          8 &         30 &         13 &         31 \cr
	
	2008 &       -1,9 &         16 &         -5 &       17,6 &       -0,2 &         20 &          2 &       22,4 &          6 &         32 &         13 &         33 \cr
	
	2007 &         -1 &         19 &       -2,7 &       16,5 &         -1 &       19,5 &          4 &         26 &          9 &       31,6 &       12,6 &       34,4 \cr
	
	2006 &       -5,5 &         18 &         -4 &         18 &         -3 &         18 &          2 &         25 &        6,5 &         28 &          8 &         34 \cr
	
	2005 &         -5 &         15 &         -5 &         14 &         -6 &       19,5 &          2 &         25 &        7,6 &       28,5 &          7 &         33 \cr
	
	2004 &         -4 &         20 &        0,8 &       15,4 &       -3,4 &       20,2 &          3 &         21 &        5,8 &       24,5 &         12 &         30 \cr
	
	2003 &         -4 &       16,4 &       -4,2 &         15 &        0,4 &       20,4 &         -5 &       23,4 &          8 &         31 &       14,6 &       34,6 \cr
	
	2002 &         -6 &       17,4 &         -1 &       16,8 &          1 &       21,6 &        0,8 &         24 &          5,60 &       26,8 &          11,40 &       37,6 \cr
	
	2001 &         -2 &         17 &         -3 &       17,8 &        1,8 &         22 &         -2 &       23,4 &        9,6 &         28 &          8 &         33 \cr
	
	2000 &         -6 &       14,3 &       -2,2 &       18,2 &       -2,6 &       19,8 &          2 &         25 &        9,6 &       28,6 &         12 &       31,4 \cr
	
	1999 &       -4,2 &       14,8 &         -6 &         15 &         -2 &       22,2 &          4 &         23 &          8 &         29 &         10 &       29,4 \cr
	
	1998 &         -4 &         16 &         -4 &       18,4 &         -2 &         20 &        4,6 &         24 &        6,8 &         29 &       10,8 &         33 \cr
	
	1997 &       -6,8 &         16 &       -2,2 &         19 &         -1 &         21 &       -1,6 &       21,4 &          7 &         29 &       10,1 &         31 \cr
	
	1996 &         -4 &         16 &       -3,8 &         15 &       -2,6 &         18 &        3,8 &         23 &          7 &       25,6 &         10 &         30 \cr
	
	1995 &         -5 &       16,6 &       -3,2 &       15,4 &       -1,2 &         17 &       -0,6 &         22 &          7 &         25 &          8 &       31,4 \cr
	
	1994 &       -2,2 &         20 &       -3,6 &       16,8 &          1 &         20 &        0,6 &         25 &          6,80 &         29 &          8,70 &       31,2 \cr
	
	1993 &         -7 &       15,4 &         -8 &       17,6 &         -4 &       18,8 &        3,8 &       25,6 &        7,5 &       28,6 &       12,6 &         31 \cr
	
	1992 &       -4,2 &       17,6 &         -4 &         19 &       -1,4 &       21,4 &        4,4 &       25,6 &        6,8 &       29,6 &       11,8 &         31 \cr
	
	1991 &         -6 &         17 &         -5 &       19,4 &          1 &       23,6 &        2,2 &         24 &          4 &         28 &          9 &         31 \cr
	
	1990 &         -5 &       16,6 &       -2,2 &         21 &         -3 &       23,2 &        2,6 &       24,2 &          6 &       27,2 &        8,2 &         35 \cr
	\hline
\end{tabular}  
\end{table}
\begin{table}
	\normalsize
	\caption{Comparison of temperature on decadal scale from July to December} 
	\normalsize
	\begin{tabular}{|c|c|c|c|c|c|c|c|c|c|c|c|c|}
		\hline
		\multicolumn{ 1}{|c}{Year} & \multicolumn{ 2}{c}{Jul} & \multicolumn{ 2}{c}{Aug} & \multicolumn{ 2}{c}{Sep} & \multicolumn{ 2}{c}{Oct} & \multicolumn{ 2}{c}{Nov} & \multicolumn{ 2}{c|}{Dec} \\
		
		\multicolumn{ 1}{|c|}{} &        Min &        Max &        Min &        Max &        Min &        Max &        Min &        Max &        Min &        Max &        Min &        Max \\
		\hline
      2014 &          14 &         32 &         13 &         30 &         10 &         28 &          5 &         26 &          6 &         21 &         -4 &         18 \cr
2013 &          13 &         35 &         15 &         34 &         12 &         32 &          9 &         25 &         -2 &         22 &         -2 &         15 \cr
2012 &         15 &         35 &         14 &         36 &         11 &         31 &          5 &         26 &          5 &         19 &         -3 &         17 \cr
2011 &         15 &         34 &         15 &         38 &         11 &         32 &          5 &         28 &         -1 &         21 &         -2 &         19 \cr    
2010 &         15 &         34 &         12 &         32 &          9 &         29 &          3 &         25 &          1 &         20 &         -4 &         17 \cr
2009 &         14 &         35 &         17 &         37 &         13 &         31 &          2 &         27 &          2 &         20 &         -6 &         16 \cr
2008 &         13 &         32 &         15 &         32 &          7 &         33 &          7 &         26 &          0 &         20 &          0 &         16 \cr
2007 &         12 &         34 &         13 &       34,6 &          9 &         29 &        1,8 &       27,4 &       -1,4 &         21 &         -4 &         16 \cr
2006 &         16 &         35 &         13 &         32 &         12 &       31,5 &        6,5 &         26 &         -1 &         19 &         -3 &         18 \cr
2005 &         15 &         39 &         11 &       33,5 &         11 &         34 &        6,5 &       24,5 &          7 &       21,5 &         -5 &         16 \cr
2004 &         12 &         33 &         15 &         33 &          9 &         34 &         11 &         25 &         -1 &         24 &         -3 &       17,5 \cr
2003 &         16 &       34,6 &       16,2 &       38,2 &         10 &         32 &        2,8 &         27 &        3,8 &       20,2 &       -4,4 &       18,2 \cr
2002 &         14 &         33 &         14 &       31,8 &          6 &       29,2 &        7,6 &       25,2 &        2,8 &         22 &        0,2 &         16 \cr
2001 &       14,8 &         36 &         15 &       36,4 &          8 &         27 &          9 &         27 &         -3 &         20 &         -5 &         16 \cr
2000 &       10,6 &         34 &       14,4 &         36 &       11,6 &         30 &        7,8 &       26,8 &          2 &       20,2 &       -1,8 &       18,4 \cr
1999 &       14,6 &         35 &       15,6 &         35 &         13 &         32 &        5,6 &       26,4 &       -6,8 &         23 &         -5 &         16 \cr
1998 &         11 &         35 &         12 &         35 &          9 &         31 &          6 &         22 &       -3,2 &         23 &         -5 &         15 \cr
1997 &         13 &         34 &         14 &         35 &          9 &         32 &          2 &         28 &          2 &         20 &         -4 &         16 \cr
1996 &         11 &         33 &         14 &         32 &          8 &         26 &        4,8 &       23,4 &          1 &         21 &         -6 &       15,4 \cr
1995 &         14 &         36 &       10,2 &         32 &          6 &         28 &          3 &       26,2 &       -2,6 &         20 &         -6 &       17,6 \cr
1994 &       15,4 &       34,1 &       16,8 &         35 &          8 &       30,2 &          3 &       26,3 &          4 &       22,2 &       -2,4 &         18 \cr
1993 &         11 &       35,2 &         13 &         36 &        9,6 &       31,7 &          6 &       24,4 &        0,4 &         20 &       -3,7 &         17 \cr
1992 &       13,2 &         35 &       16,6 &       36,8 &          9 &         31 &        6,7 &       23,6 &       -0,8 &       21,2 &         -2 &         17 \cr
1991 &         13 &         35 &         12 &         34 &         14 &       31,6 &        2,8 &       26,2 &        2,8 &         20 &       -5,2 &       13,6 \cr
1990 &         14 &         34 &         14 &         36 &          9 &         29 &          9 &       30,2 &       -2,6 &         20 &         -3 &         14 \cr
		\hline
	\end{tabular}  
\end{table}
\centering
\begin{table}
\centering \scriptsize
\caption{Day Time and Night Time Variation} 
\begin{tabular}{|c|c|c|c|c|c|c|c|c|c|c|c|c|}
\hline
\bf Year & \bf Jan &\bf Feb &\bf Mar & \bf Apr &\bf May&\bf Jun & \bf Jul &\bf Aug&\bf Sep&\bf Oct &\bf Nov&\bf Dec \cr
\hline
\bf 2014 &  7 &  6 & 10 &  8 & 10 & 11 &  8 &  9 & 10 &  9 &  6 &  7 \cr \hline
\bf 2013 &  7 &  8 &  7 &  9 &  8 & 11 & 11 & 12 &  9 &  7 &  8 &  9 \cr \hline
\bf 2012 & 10 &  9 & 13 &  9 & 11 & 11 & 12 & 12 & 11 &  9 &  8 &  7 \cr \hline
\bf 2011 &  7 &  8 &  9 & 10 & 12 &  9 & 10 & 12 & 11 & 10 & 11 &  9 \cr \hline
\bf 2010 &  7 &  8 &  8 &  9 & 10 & 10 & 11 &  9 & 10 &  8 &  6 &  6 \cr \hline
\bf 2009 &  8 &  9 &  9 &  8 & 12 & 10 & 12 & 12 & 11 & 10 &  9 &  7 \cr \hline
\bf 2008 &  8 & 10 &  9 & 11 & 11 & 11 & 11 & 11 & 11 &  9 &  7 &  7 \cr \hline
\bf 2007 &  8 &  9 &  9 & 13 & 11 & 10 & 13 & 11 & 12 & 11 & 10 & 10 \cr \hline
\bf 2006 & 10 & 10 & 10 & 12 & 12 & 14 & 13 & 11 & 10 & 11 &  9 &  9 \cr \hline
\bf 2005 &  9 &  9 &  9 & 10 & 11 & 12 & 12 & 12 & 10 &  9 &  9 &  8 \cr \hline
\bf 2004 &  8 &  7 &  9 &  9 & 10 & 10 & 12 & 10 & 12 &  7 &  9 &  7 \cr \hline
\bf 2003 &  8 & 12 & 12 & 11 & 13 & 12 & 11 & 11 & 11 &  9 &  7 &  8 \cr \hline
\bf 2002 & 11 &  9 & 12 & 10 & 10 & 12 & 11 & 10 &  9 &  9 &  6 &  6 \cr \hline
\bf 2001 &  7 & 10 &  7 & 11 & 11 & 13 & 11 & 12 & 11 & 10 &  8 &  8 \cr \hline
\bf 2000 &  9 & 12 & 10 &  9 & 11 & 13 & 12 & 13 & 12 &  9 &  6 &  7 \cr \hline
\bf 1999 &  9 & 11 & 11 & 10 & 11 & 12 & 13 & 11 & 12 &  9 & 10 &  8 \cr \hline
\bf 1998 &  8 & 12 & 11 &  9 & 12 & 12 & 13 & 11 & 10 &  9 &  8 &  7 \cr \hline
\bf 1997 &  9 &  9 & 12 & 11 & 12 & 10 & 12 & 11 & 12 &  9 &  7 &  8 \cr \hline
\bf 1996 &  7 &  9 & 10 & 10 & 10 & 12 & 12 & 12 & 10 & 10 &  7 &  8 \cr \hline
\bf 1995 & 10 &  9 &  9 & 10 &  9 & 11 & 12 & 11 & 10 & 12 &  9 &  8 \cr \hline
\bf 1994 &  9 &  9 & 11 & 10 & 10 & 11 & 13 & 11 &  9 & 10 &  8 &  7 \cr \hline
\bf 1993 &  6 &  6 &  8 & 13 & 17 & 21 & 22 & 24 & 20 & 16 & 10 &  9 \cr \hline
\bf 1992 & 10 & 11 & 11 & 10 & 13 & 10 & 12 & 12 & 11 &  7 &  8 &  8 \cr \hline
\bf 1991 & 11 &  9 & 11 & 12 & 11 & 12 & 13 & 12 & 11 &  9 &  7 & 12 \cr \hline
\bf 1990 & 13 & 11 & 12 & 11 & 12 & 12 & 12 & 13 & 11 &  9 & 10 &  9 \cr \hline

\bf Average 1990-2014&8,49 &9,36  &9,98 &10,15 &11,27 &11,55 &12,15 &11,78 &11,07 &9,38 &8,13 &7,90 \cr  \hline

\end{tabular}
\end{table}


\begin{table}
\tiny
\centering
\caption{Rain/Year [mm]}

\begin{tabular}{|c|c|c|c|c|c|c|c|c|c|c|c|c|c|c|c|}
\hline
\bf Year & & Jan & Feb & Mar & Apr & May & Jun & Jul & Ago & Sep & Oct & Nov &  Dec & \bf Tot & \bf \%Year \\
\hline
&  &  &  & &  & & & & & & & & & &          \cr
\hline
      2014 & \bf  62 &  45,50 & 183,10 &  21,10 &  26,60 &   6,20 &  20,40 & 177,40 &  18,00 &  43,50 & 205,10 &  74,70 &  15,20 & \bf 836,80 & \bf 3,19\% \cr \hline
2013 & \bf 152 & 119,70 & 181,30 & 575,50 & 172,30 &  61,80 &   7,00 &  53,50 &  46,90 & 124,80 &  76,10 & 260,10 &  34,80 & \bf 1713,80 & \bf 6,53\% \cr \hline
2012 & \bf  39 &   9,30 &  36,10 &  14,80 & 127,40 &  61,60 &  27,30 &  17,20 &  19,00 & 173,60 &  83,80 & 239,70 & 119,00 & \bf 928,80 & \bf 3,54\% \cr \hline
2011 & \bf  52 &  32,70 &  57,00 &  75,80 &  76,70 &  48,50 &  97,30 &  14,50 &  14,10 &  68,00 &  95,20 &  53,60 & 171,70 & \bf 805,10 & \bf 3,07\% \cr \hline
2010 & \bf  59 &  51,30 &  73,00 &  51,90 &  41,60 & 155,30 &  65,80 &  13,80 &  15,80 &  48,00 &  24,60 & 228,90 & 143,30 & \bf 913,30 & \bf 3,48\% \cr \hline
2009 & \bf  57 &  74,10 &  81,40 & 113,60 &  66,10 &  22,00 &  79,50 &   1,00 &  15,80 &  42,70 &  51,40 &  51,60 &  19,00 & \bf 618,20 & \bf 2,35\% \cr \hline
2008 & \bf  67 & 150,00 &  31,60 &  96,50 &  62,90 & 118,60 &  13,40 &   1,00 &  24,00 &  17,80 & 185,40 &  56,60 &   3,00 & \bf 760,80 & \bf 2,90\% \cr \hline
2007 & \bf  68 &  45,30 &  50,70 &  76,90 & 134,50 &  62,80 &   4,00 &   6,60 &  73,70 &  60,50 & 103,40 & 113,90 &  18,30 & \bf 750,60 & \bf 2,86\% \cr \hline
2006 & \bf  80 &  84,40 &  99,90 &  38,60 &  72,20 &  55,60 &  57,30 &  73,00 &  45,90 & 148,90 & 116,60 & 160,30 &   3,00 & \bf 955,70 & \bf 3,64\% \cr \hline
2005 & \bf  48 &  36,80 &  79,40 &  40,70 &  53,30 &  54,10 &  36,90 &  13,30 &  38,10 &  78,90 &  81,90 &  13,20 &  51,20 & \bf 577,80 & \bf 2,20\% \cr \hline
2004 & \bf  97 & 151,60 & 118,50 & 106,00 & 157,60 &  96,20 &  28,20 &  36,50 &  83,70 &  88,30 & 106,00 & 138,60 &  53,10 & \bf 1164,30 & \bf 4,44\% \cr \hline
2003 & \bf  77 &  98,80 &  28,00 &  30,90 &  70,30 &  36,30 &  36,80 &  24,40 &  16,90 &  40,00 & 297,60 & 167,80 &  72,80 & \bf 920,60 & \bf 3,51\% \cr \hline
2002 & \bf 101 &  38,70 &  69,80 &   6,20 &  69,30 &  64,28 &  64,90 &  70,00 & 151,20 & 212,70 & 136,60 & 150,70 & 178,90 & \bf 1213,28 & \bf 4,62\% \cr \hline
2001 & \bf  64 &  77,20 &  29,40 & 108,70 &  65,50 &  39,70 &  18,90 &  29,70 &   3,00 & 143,90 &  78,70 & 112,00 &  62,00 & \bf 768,70 & \bf 2,93\% \cr \hline
2000 & \bf 100 &  70,30 &  10,40 & 106,70 & 102,30 &  24,80 &  38,30 &  41,50 &  20,60 &  71,40 & 159,10 & 443,70 & 105,60 & \bf 1194,70 & \bf 4,55\% \cr \hline
1999 & \bf  75 &  52,10 &  41,20 &  49,60 &  68,90 &  22,40 &  71,40 &  11,00 &  25,70 & 149,30 & 153,70 & 137,50 & 114,10 & \bf 896,90 & \bf 3,42\% \cr \hline
1998 & \bf  67 &  60,60 & 109,40 &  50,70 &  82,80 &  52,70 &  49,50 &  23,50 &  20,70 & 118,10 & 125,70 &  56,10 &  54,90 & \bf 804,70 & \bf 3,07\% \cr \hline
1997 & \bf  76 & 120,20 & 189,10 &  29,60 &  80,60 &  45,60 &  94,30 &  21,10 &  60,40 &  23,30 &  40,60 &  95,80 & 114,30 & \bf 914,90 & \bf 3,49\% \cr \hline
1996 & \bf 102 & 121,30 &  75,50 &  28,90 &  78,40 &  70,00 &  32,90 &  10,90 &  89,10 & 120,20 & 206,00 & 271,50 & 116,20 & \bf 1220,90 & \bf 4,65\% \cr \hline
1995 & \bf  89 &  56,60 & 129,00 &  87,90 &  65,10 & 130,80 &  53,20 &   8,70 & 134,70 & 103,30 &  94,40 &  66,30 & 132,70 & \bf 1062,70 & \bf 4,05\% \cr \hline
1994 & \bf 126 & 163,50 &  10,30 & 266,20 & 126,10 &  39,00 &  82,00 &   6,00 & 127,50 & 432,70 &  89,00 &  81,50 &  86,50 & \bf 1510,30 & \bf 5,75\% \cr \hline
1993 & \bf 102 &   3,00 &   2,00 &  39,70 &  85,50 &  16,70 &   5,50 &   2,50 & 282,60 & 266,50 & 229,30 & 226,00 &  61,40 & \bf 1220,70 & \bf 4,65\% \cr \hline
1992 & \bf  84 &  21,20 &   8,20 &  17,10 &  84,80 & 157,60 & 101,90 &  26,60 &  72,60 &  62,20 & 295,40 &  86,10 &  73,80 & \bf 1007,50 & \bf 3,84\% \cr \hline
1991 & \bf 137 &  22,10 &  55,70 &  92,60 &  66,40 & 176,30 & 109,60 & 166,10 &  25,00 & 425,00 & 249,80 & 247,60 &  10,40 & \bf 1646,60 & \bf 6,27\% \cr \hline
1990 & \bf 154 &  45,90 &  71,90 &  39,40 & 245,20 &  25,20 &  46,50 &  28,40 &  54,20 & 223,10 & 637,00 & 228,20 &  199,80 & \bf 1844,80 & \bf 7,03\% \cr \hline
& & & & & & & & & & & & & & & \\
\bf Average 1990-2014 & \bf 85,31 & \bf 70,09 & \bf 72,88 & \bf 86,62 & \bf 91,30 & \bf 65,76 & \bf 49,71 & \bf 35,13 & \bf 59,17 & \bf 131,47 & \bf 156,90 & \bf 150,48 & \bf 80,60 & \bf 26252,48 & \cr
\hline
\bf &&&&&&&&&&&&& \bf Year Average & \bf 1050,099 & \\ \hline
\end{tabular}
\end{table}
\begin{table}
\caption{Wind Average[1990-2014]/Year on Km/h}
\tiny
\begin{tabular}{|c|c|c|c|c|c|c|c|c|c|c|c|c|c|c|c|} 
\hline
\multicolumn{ 14}{|c}{} &            &            \\
\hline
\multicolumn{ 1}{|c|}{Day} &        Jan &        Feb &        Mar &        Apr &        May &        Jun &        Jul &        Ago &        Sep &        Oct &        Nov &        Dec &\bf Average &  \bf Min &  \bf Max  \\
\hline
\multicolumn{ 1}{|c|}{} &            &            &            &            &            &            &            &            &            &            &            &            &            &            &            \\

1 & 21,08 & 21,63 &  26,08 & 23,60 & 18,92 & 19,21 & 22,64 & 21,64 & 22,54 &  18,25 & 25,88 & 22,32 & \bf 21,98 &  18,25 & 26,08 \\

2 & 20,44 & 20,00 & 22,75 & 23,80 & 19,12 & 18,87 & 23,92 & 20,04 & 22,25 & 22,25 & 25,52 & 21,96 & {\bf 21,74} & 18,87 & 25,52 \\

3 & 18,64 & 19,58 & 22,33 & 21,75 & 22,24 & 20,25 & 23,32 & 20,32 & 21,00 &  26,33 & 19,12 & 23,28 & \bf 21,51 & 18,64 & 26,33 \\

4 &      22,29 &      21,33 &      23,83 &      22,25 &      22,40 &      22,96 &      22,20 &      19,84 &      25,04 &      19,96 &      22,72 &      22,21 & {\bf 22,25} &      19,84 &      25,04 \\

5 &      21,58 &      21,25 &      24,63 &      21,83 &      20,17 &      21,54 &      23,76 &      21,44 &      22,20 &      21,13 &      24,16 &      23,24 & {\bf 22,24} &      20,17 &      24,63 \\

6 &      21,35 &      20,83 &      22,33 &      24,25 &      20,21 &      22,00 &      20,68 &      21,60 &      19,56 &      18,83 &      24,64 &      20,42 & {\bf 21,39} &      18,83 &      24,64 \\

7 &      18,96 &      20,21 &      22,92 &      27,30 &      20,92 &      21,21 &      22,56 &      20,44 &      19,96 &      20,79 &      23,88 &      18,67 & {\bf 21,48} &      18,67 &      27,30 \\

8 &      18,75 &      23,00 &      24,00 &      20,92 &      19,63 &      21,00 &      25,48 &      20,76 &      21,44 &      20,08 &      22,80 &      21,04 & {\bf 21,58} &      18,75 &      25,48 \\

9 &      17,04 &      24,71 &      24,56 &      23,08 &      21,08 &      20,33 &      21,48 &      20,68 &      22,64 &      20,21 &      19,76 &      26,04 & {\bf 21,80} &      17,04 &      26,04 \\

10 &      15,83 &      20,50 &      22,04 &      22,64 &      18,50 &      21,71 &      23,36 &      21,36 &      21,60 &      21,04 &      20,16 &      23,96 & {\bf 21,06} &      15,83 &      23,96 \\

11 &      19,17 &      21,63 &      19,60 &      25,58 &      18,50 &      21,79 &      24,16 &      23,48 &      20,80 &      19,50 &      23,96 &      20,52 & {\bf 21,56} &      18,50 &      25,58 \\

12 &      20,17 &      19,79 &      21,00 &      25,25 &      19,96 &      21,52 &      21,84 &      22,76 &      23,52 &      22,60 &      23,08 &      17,76 & {\bf 21,60} &      17,76 &      25,25 \\

13 &      20,46 &      24,83 &      21,08 &      23,29 &      22,08 &      20,52 &      21,12 &      20,60 &      26,16 &      21,56 &      21,75 &      17,76 & {\bf 21,77} &      17,76 &      26,16 \\

14 &      19,63 &      19,50 &      18,00 &      23,96 &      21,08 &      20,72 &      22,20 &      21,04 &      26,12 &      20,32 &      21,83 &      19,52 & {\bf 21,16} &      18,00 &      26,12 \\

15 &      22,96 &      20,44 &      18,56 &      22,42 &      19,17 &      21,20 &      22,20 &      19,92 &      23,52 &      21,68 &      20,83 &      28,60 & {\bf 21,79} &      18,56 &      28,60 \\

16 &      19,54 &      24,75 &      24,28 &      24,67 &      20,83 &      20,24 &      20,76 &      20,12 &      19,32 &      21,08 &      23,75 &      27,68 & {\bf 22,25} &      19,32 &      27,68 \\

17 &      18,21 &      21,79 &      22,04 &      23,58 &      18,54 &      20,92 &      20,40 &      21,24 &      18,16 &      24,08 &      24,29 &      25,52 & {\bf 21,56} &      18,16 &      25,52 \\

18 &      17,71 &      21,54 &      21,88 &      23,48 &      20,29 &      20,78 &      21,96 &      19,52 &      19,00 &      25,08 &      23,92 &      20,44 & {\bf 21,30} &      17,71 &      25,08 \\

19 &      20,38 &      20,79 &      21,56 &      23,20 &      22,96 &      21,13 &      21,28 &      20,16 &      18,72 &      23,64 &      24,79 &      18,64 & {\bf 21,44} &      18,64 &      24,79 \\

20 &      18,96 &      20,83 &      23,56 &      22,83 &      22,21 &      22,64 &      21,56 &      21,00 &      20,64 &      20,36 &      23,42 &      21,60 & {\bf 21,63} &      18,96 &      23,56 \\

21 &      17,71 &      20,88 &      22,52 &      23,38 &      22,54 &      20,35 &      19,48 &      21,04 &      20,40 &      22,16 &      24,33 &      22,21 & {\bf 21,42} &      17,71 &      24,33 \\

22 &      22,63 &      22,33 &      22,24 &      19,96 &      20,38 &      21,67 &      20,40 &      20,16 &      20,08 &      20,28 &      23,21 &      21,59 & {\bf 21,24} &      19,96 &      23,21 \\

23 &      21,96 &      21,96 &      20,28 &      21,92 &      20,17 &      19,75 &      20,96 &      19,64 &      22,68 &      21,88 &      22,67 &      23,04 & {\bf 21,41} &      19,64 &      23,04 \\

24 &      20,17 &      21,25 &      23,96 &      20,72 &      19,70 &      18,17 &      22,20 &      19,88 &      22,33 &      20,04 &      21,04 &      23,70 & {\bf 21,10} &      18,17 &      23,96 \\

25 &      21,33 &      20,17 &      22,46 &      21,24 &      19,92 &      18,71 &      20,68 &      22,88 &      23,32 &      20,16 &      18,38 &      23,22 & {\bf 21,04} &      18,38 &      23,32 \\

26 &      24,13 &      22,92 &      22,64 &      18,84 &      19,26 &      20,21 &      21,08 &      20,56 &      23,64 &      21,72 &      19,25 &      25,00 & {\bf 21,60} &      18,84 &      25,00 \\

27 &      22,00 &      23,87 &      25,20 &      20,84 &      20,91 &      20,61 &      21,28 &      20,88 &      21,52 &      22,08 &      18,58 &      24,57 & {\bf 21,86} &      18,58 &      25,20 \\

28 &      20,50 &      22,71 &      22,32 &      19,76 &      19,87 &      22,71 &      20,24 &      23,13 &      20,38 &      21,40 &      19,04 &      23,74 & {\bf 21,32} &      19,04 &      23,74 \\

29 &      20,88 &      26,29 &      24,16 &      20,04 &      19,48 &      21,88 &      22,36 &      22,88 &      19,54 &      23,00 &      18,88 &      19,48 & {\bf 21,57} &      18,88 &      26,29 \\

30 &      21,83 &            &      20,84 &      21,36 &      19,41 &      21,46 &      21,75 &      23,76 &      18,50 &      24,40 &      20,00 &      21,13 & {\bf 21,31} &      18,50 &      24,40 \\

31 &      22,88 &            &      22,44 &            &      19,30 &            &      22,13 &      24,20 &            &      21,60 &            &      19,70 & {\bf 21,75} &      19,30 &      24,20 \\
\hline
{\bf Average} & {\bf 20,29} & {\bf 21,77} & {\bf 22,45} & {\bf 22,59} & {\bf 20,31} & {\bf 20,87} & {\bf 21,92} & {\bf 21,19} & {\bf 21,55} & {\bf 21,53} & {\bf 22,19} & {\bf 22,21} & {\bf Average} & {\bf 15,83} & {\bf 28,60} \\

{\bf Min} &      15,83 &      19,50 &      18,00 &      18,84 &      18,50 &      18,17 &      19,48 &      19,52 &      18,16 &      18,25 &      18,38 &      17,76 & {\bf 21,04} &            &            \\

{\bf Max} &      24,13 &      26,29 &      26,08 &      27,30 &      22,96 &      22,96 &      25,48 &      24,20 &      26,16 &      26,33 &      25,88 &      28,60 & {\bf 22,25} &            &            \\

\hline
\end{tabular}  
\end{table}
\centering
\begin{table}
\centering \scriptsize
\caption{Trend of Temperature of Pieve a Nievole in Jan-Feb-Mar and Jun-Jul-Aug} 
\begin{tabular}{|c|c|c|c|c|c|c|c|c|}
\hline
\bf Year & \bf Jan & \bf Feb &\bf Mar&\bf Jun & \bf Jul &\bf Aug \cr
\hline
{\bf 2014} & 9,03 & 10,88 & 11,27 & 21,45 &  21,56 & 21,92 \\
{\bf 2013} &       6,85 &       5,45 &       9,40 &      19,27 &      23,42 &      24,37 \\
{\bf 2012} &       6,11 &       1,55 &      11,00 &      21,50 &      24,03 &      25,31 \\
{\bf 2011} &       7,08 &       7,54 &       9,56 &      21,60 &      22,98 &      24,34 \\
{\bf 2010} &       5,13 &       7,55 &       9,50 &      20,40 &      25,04 &      23,00 \\
{\bf 2009} &       6,98 &       6,55 &       9,85 &      21,10 &      23,85 &      25,60 \\
{\bf 2008} &       8,45 &       8,19 &       9,95 &      21,05 &      23,58 &      24,05 \\
{\bf 2007} &       9,08 &       9,28 &      10,42 &      21,17 &      22,60 &      22,52 \\
{\bf 2006} &       5,14 &       7,08 &       8,96 &      20,80 &      26,24 &      22,50 \\
{\bf 2005} &       5,69 &       4,67 &       8,89 &      21,90 &      24,27 &      23,68 \\
{\bf 2004} &       6,11 &       8,74 &       9,42 &      20,51 &      22,74 &      24,42 \\
{\bf 2003} &       6,51 &       4,91 &      10,07 &      25,61 &      26,58 &      27,75 \\
{\bf 2002} &       5,70 &       7,90 &      11,30 &      18,80 &      23,50 &      22,90 \\
{\bf 2001} &       8,65 &       8,54 &      13,10 &      20,46 &      23,95 &      25,58 \\
{\bf 2000} &       5,76 &       7,92 &      10,45 &      22,12 &      22,12 &      24,23 \\
{\bf 1999} &       5,82 &       5,50 &      10,40 &      20,04 &      23,90 &      25,00 \\
{\bf 1998} &       7,53 &       7,92 &       9,20 &      20,79 &      23,63 &      24,89 \\
{\bf 1997} &       8,09 &       9,30 &      10,89 &      20,91 &      22,60 &      24,25 \\
{\bf 1996} &       8,61 &       6,61 &       9,43 &      21,09 &      22,49 &      22,54 \\
{\bf 1995} &       6,30 &       8,69 &       9,35 &      18,45 &      24,63 &      22,85 \\
{\bf 1994} &       8,26 &       7,35 &      11,81 &      19,34 &      24,90 &      25,51 \\
{\bf 1993} &       6,51 &       6,09 &       8,09 &      21,21 &      21,81 &      24,81 \\
{\bf 1992} &       7,56 &       7,73 &      10,70 &      20,13 &      23,31 &      25,00 \\
{\bf 1991} &       6,97 &       6,45 &      12,98 &      19,35 &      24,47 &      24,69 \\
{\bf 1990} &       6,89 &      10,74 &      10,76 &      20,55 &      23,31 &      24,05 \\
\hline
\bf Average 1990-2014 & \bf 6,99 & \bf 7,33 & \bf 10,27 & \bf 20,78 & \bf 23,66 & \bf 24,23  \cr  \hline

\end{tabular}
\end{table}
\newpage \clearpage

\begin{figure}
\begin{center}
\includegraphics[width=1\textwidth]{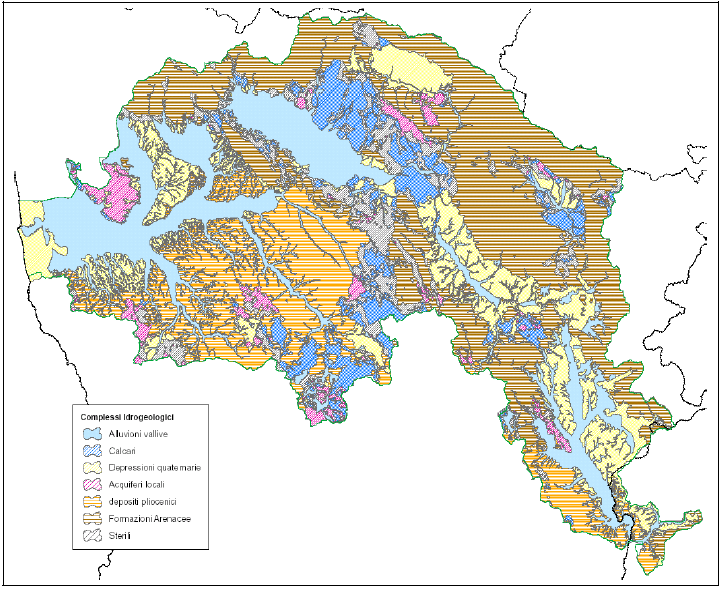} \caption{Hydro-Geological complexes Map of the Arno river basin \cite{Autarno:2006}} 
\end{center}
\end{figure}


\begin{figure}
\begin{center}
\includegraphics[width=1.3\textwidth{}]{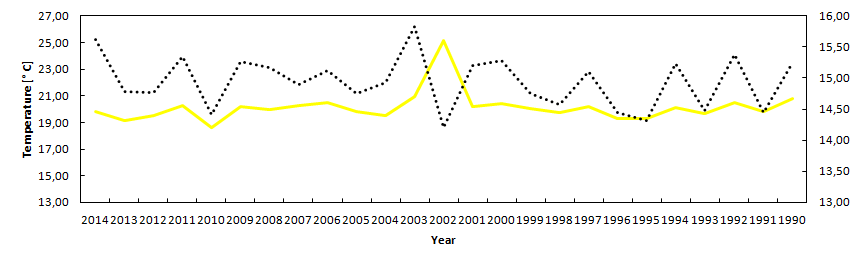}
\caption{Maximum Temperature (dotted line) and Average Temperature Variations 1990-2014 (solid line)}
\end{center}
\end{figure}
\begin{figure}
\begin{center}
\includegraphics[width=1.3\textwidth{}]{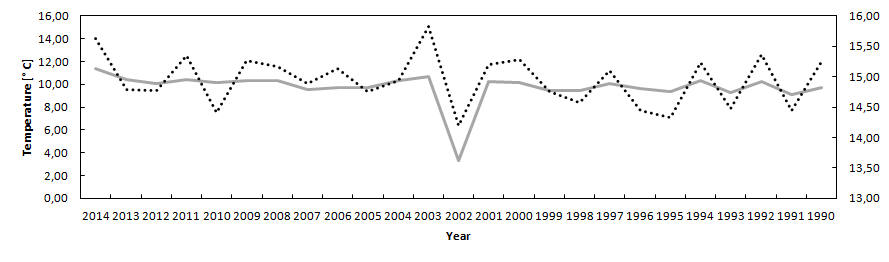}
\caption{Minimum Temperature (dotted line) and Average Temperature Variations 1990-2014 (solid line)}
\end{center}
\end{figure}
\begin{figure}
\begin{center}
\includegraphics[width=1.3\textwidth{}]{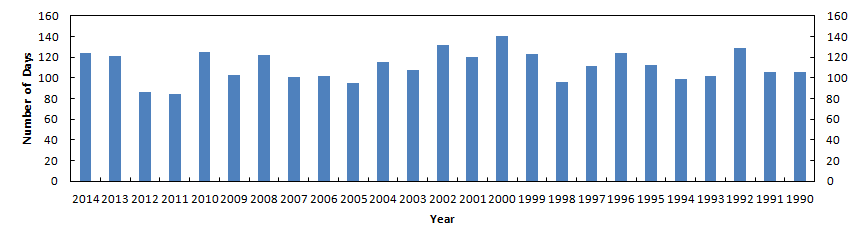}
\caption{Total Days of Rain by Year}
\end{center}
\end{figure}
\begin{figure}
\begin{center}
\includegraphics[width=1.3\textwidth{}]{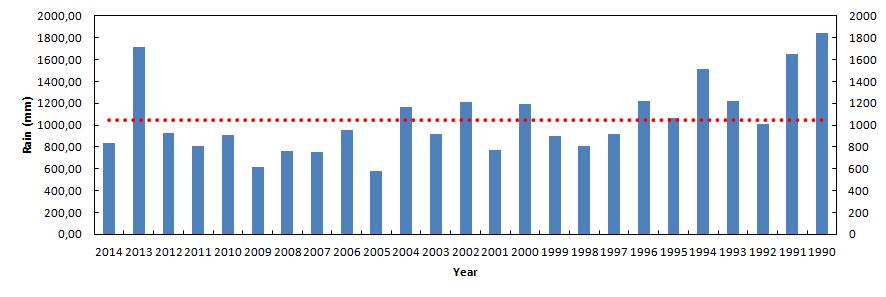}
\caption{Total rain [mm] by year}
\end{center}
\end{figure}
\begin{figure}
\begin{center}
\includegraphics[width=1.3\textwidth{}]{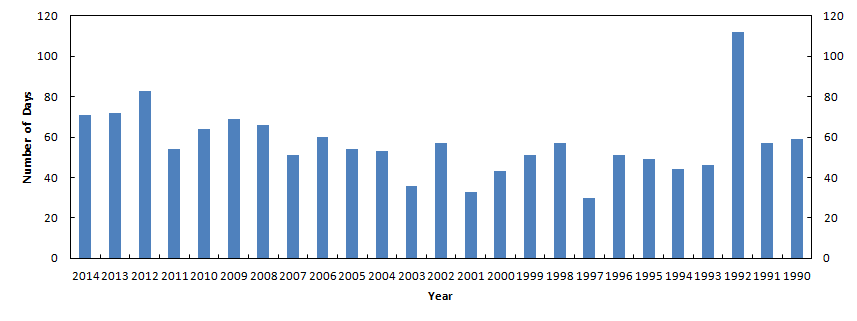}
\caption{Total day of Fog by year}
\end{center}
\end{figure}
\begin{figure}
\begin{center}
\includegraphics[width=1.3\textwidth{}]{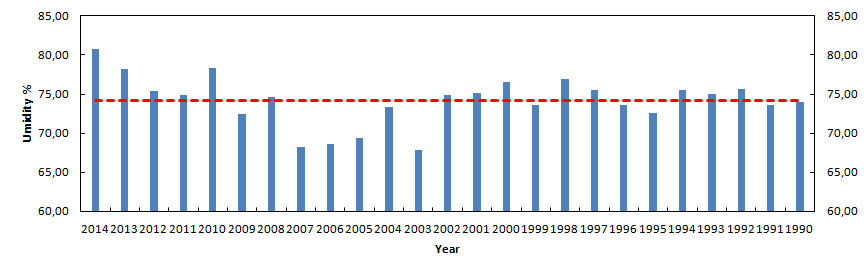}
\caption{Umidity (solid line) and Average Umidity Variations 1990-2014 (dotted line) }
\end{center}
\end{figure}
\begin{figure}
\begin{center}
\includegraphics[width=1.3\textwidth{}]{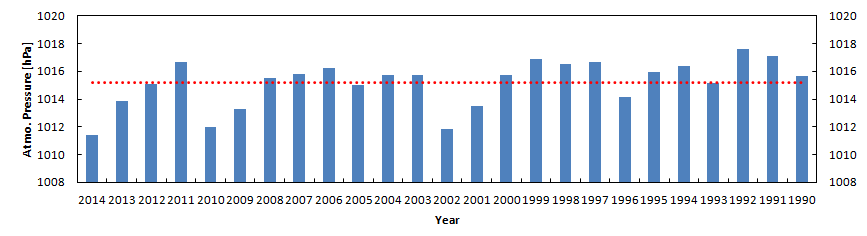}
\caption{Pressure (solid line) and Average Pressure Variations 1990-2014 (dotted line) }
\end{center}
\end{figure}
\begin{figure}
\begin{center}
\includegraphics[width=1.3\textwidth{}]{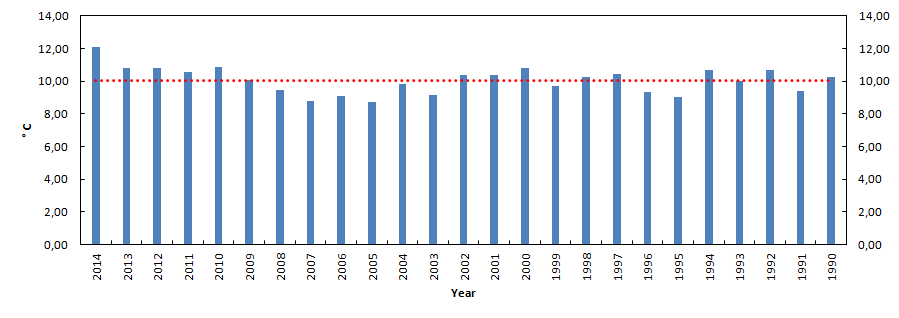}
\caption{Dew Point (solid line) and Average Dew Point Variations 1990-2014 (dotted line) }
\end{center}
\end{figure}
\begin{figure}
\begin{center}
\includegraphics[width=1.3\textwidth{}]{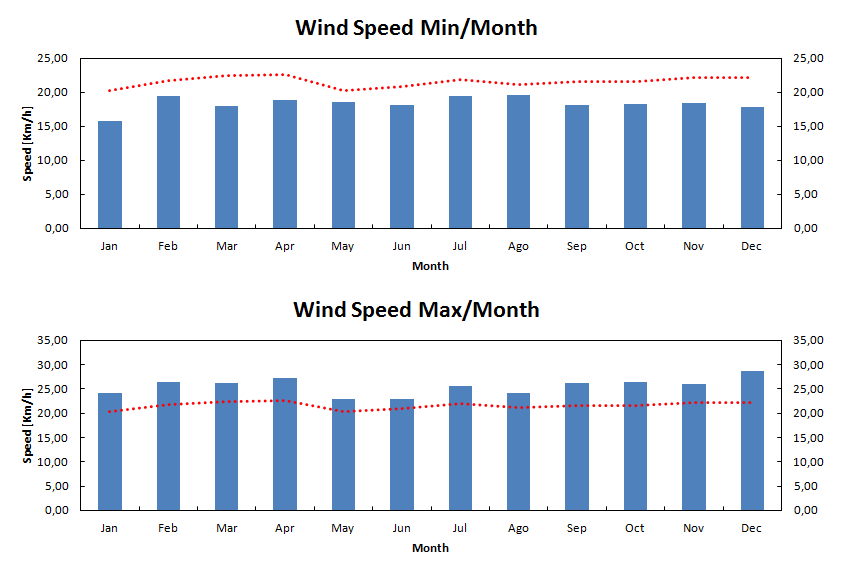}
\caption{Wind Speed (solid line) and Average Wind Speed Variations 1990-2014 (dotted line) }
\end{center}
\end{figure}
\clearpage
\begin{figure}
\includegraphics[width=1.3\textwidth{}]{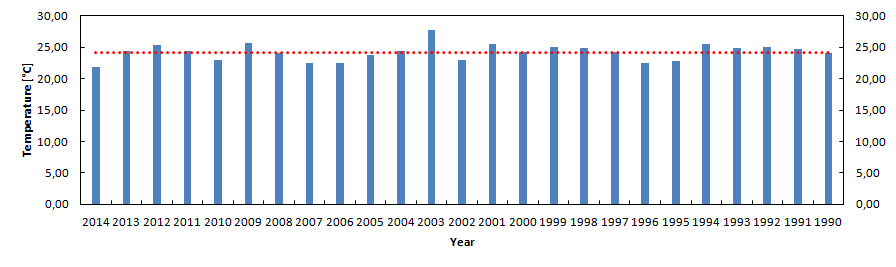}
\caption{August Temperature (solid line) and Average August Temperature 1990-2014 (dotted line) }
\end{figure}
\begin{figure}
\includegraphics[width=1.3\textwidth{}]{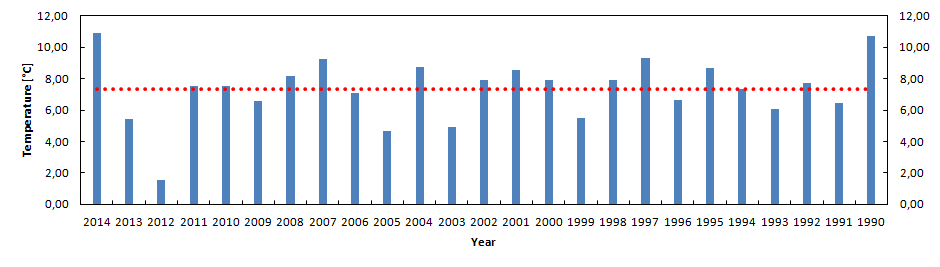}
\caption{February Temperature (solid line) and Average February Temperature 1990-2014 (dotted line)}
\end{figure}

\centering
\begin{figure}
\begin{center}
\includegraphics[width=1.3\textwidth{}]{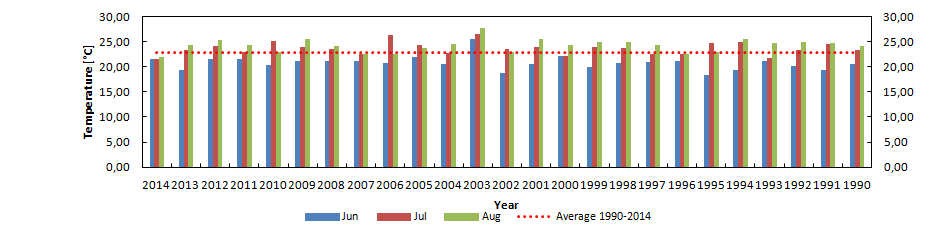}
\caption{Summer Temperature (solid line) and Average Temperature Variations 1990-2014 (dotted line). Jun: blu, Jul: red Aug: green}
\end{center}
\end{figure}
\begin{figure}
\begin{center}
\includegraphics[width=1.3\textwidth{}]{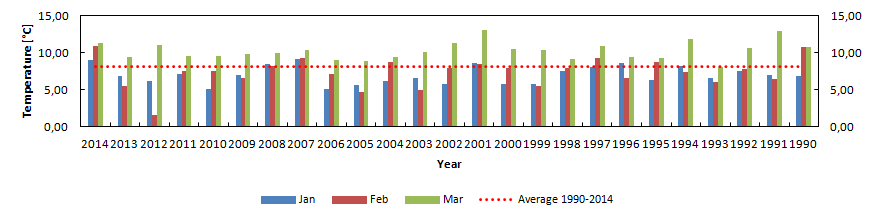}
\caption{Winner Temperature (solid line) and Average Temperature Variations 1990-2014 (dotted line) Jan: blu, Feb: red Mar: green}
\end{center}
\end{figure}
\begin{figure}
\caption{Total Rain (mm) by Month}
\includegraphics[width=0.55\textwidth{}]{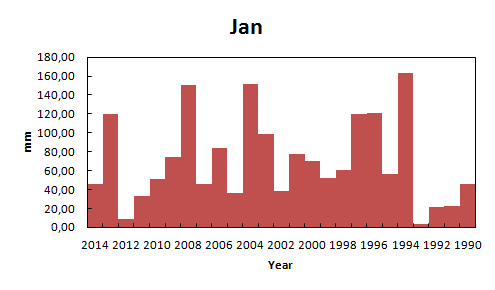} \quad 
\includegraphics[width=0.55\textwidth{}]{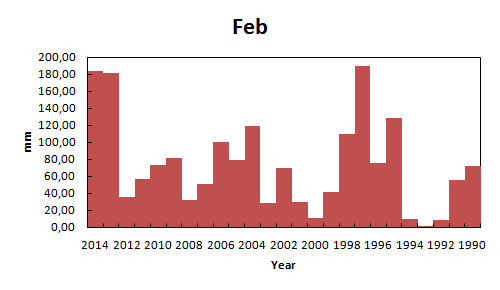} \quad
\includegraphics[width=0.55\textwidth{}]{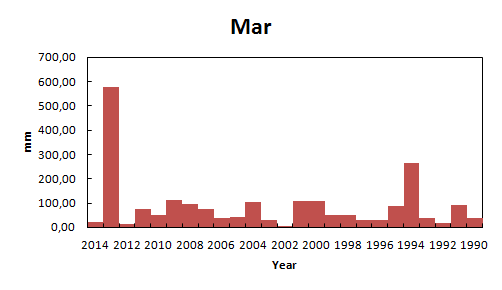} \quad
\includegraphics[width=0.60\textwidth{}]{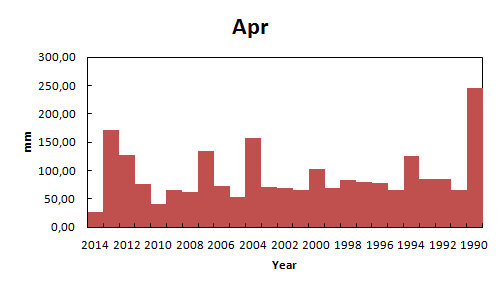} \quad
\includegraphics[width=0.55\textwidth{}]{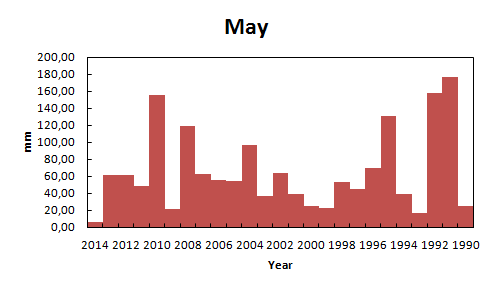} \quad 
\includegraphics[width=0.55\textwidth{}]{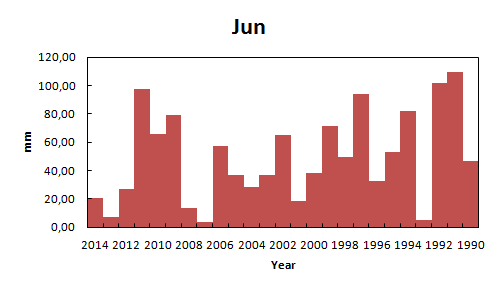} \quad
\includegraphics[width=0.55\textwidth{}]{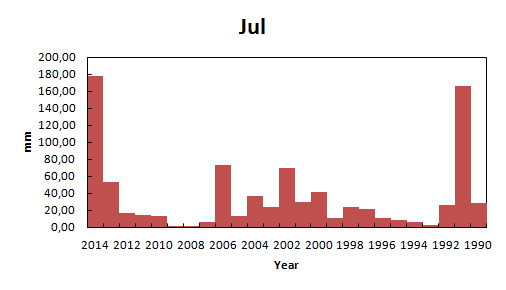} \quad
\includegraphics[width=0.55\textwidth{}]{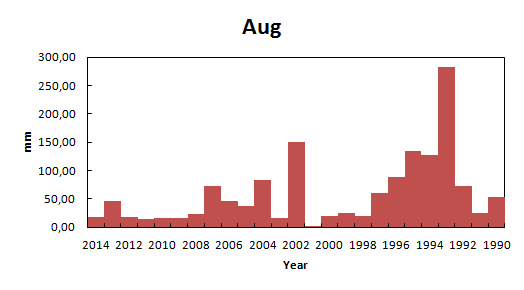} \quad
\includegraphics[width=0.55\textwidth{}]{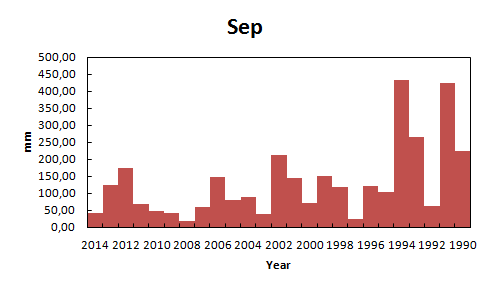} \quad 
\includegraphics[width=0.55\textwidth{}]{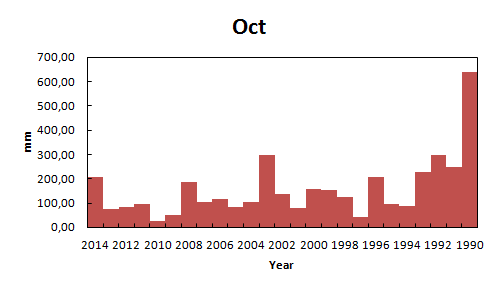} \quad
\includegraphics[width=0.55\textwidth{}]{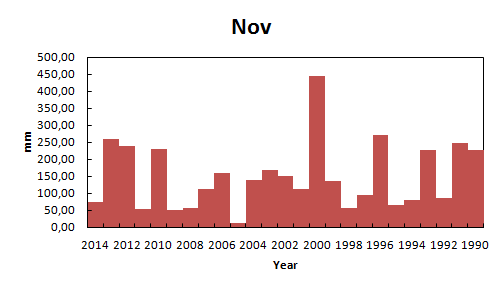} \quad
\includegraphics[width=0.55\textwidth{}]{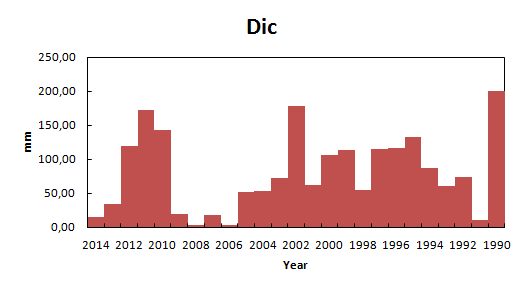} \\
\end{figure}
\begin{figure}
\caption{Solar Irradiation Map}
\includegraphics[width=1\textwidth{}]{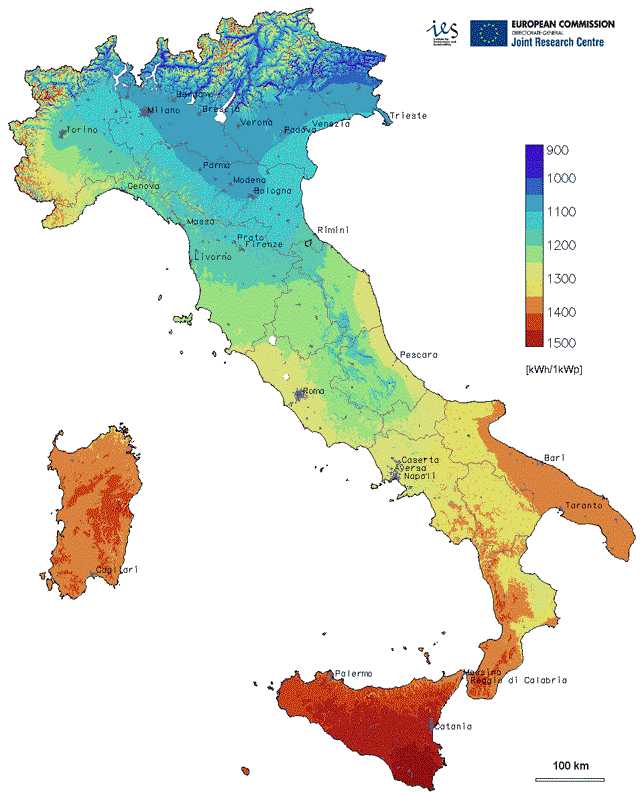}
\end{figure}
\begin{figure}
\includegraphics[width=0.9\textwidth{}]{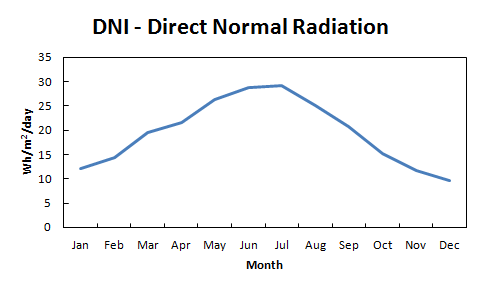}
\caption{DNI Direct Normal Radiation (Wh/m$^2$/day)} 
\end{figure}
\begin{figure}
	\includegraphics[width=0.9\textwidth{}]{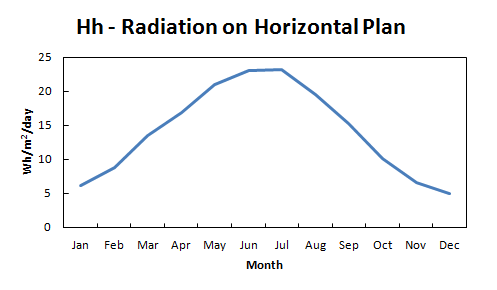}
	\caption{Hh Radiation on Hozizontal Plan (Wh/m$^2$/day)} 
\end{figure}
\clearpage 
\centering


\addcontentsline{toc}{section}{\refname}

\begin{thebibliography}{99}
\bibitem{Tasselli:2018} 
\textbf {Tasselli, D. Ricci, S., Bianchi P.}, "PAN.R.C. - A New Level 3 Biosafety and Astrobiology Laboratory in Pieve a Nievole (PT)",  arXiv:1811.05201v1 [astro-ph.IM], 2018. 
\bibitem{Tasselli:2011ug} 
\textbf {Ente per le Nuove Tecnologie, l'Energia e l'Ambiente}, ``Tabella dei gradi/giorno dei Comuni italiani raggruppati per Regione e Provincia, Legge 26 agosto 1993, n. 412, allegato A'', 1 marzo 2011, p. 151.  
\bibitem{mappageologica:2013} 
\textbf {Prof. L. Carmignani, Prof. G. Cornamusini, Dott. I. Callegari}, ``Mappa Geologica di Pieve a Nievole (PT)'', Università di Siena, Dipartimento di Scienze della Terra - Centro di GeoTecnologie, Foglio 262.
\bibitem{datimeteo:2013} 
\textbf {IlMeteo.it} ``Archivio Dati Meteo Pieve a Nievole (PT)'', IlMeteo.it.
\bibitem{eumetsat:2013} 
\textbf {Eumetsat} ``Archivio Dati Meteo Pieve a Nievole (PT)'', Eumetsat. 
\bibitem{atlantevento:2013} 
\textbf {Atlantedelvento} ``Archivio Dati Atlante Eolico'', http://www.atlanteeolico.it. 
\bibitem{Autarno:2006} 
\textbf {Autorità di Bacino Fiume Arno} ``Complessi Idrogeologici Bacino del Fiume Arno'', Autorità di Bacino Fiume Arno. 
\bibitem{Dellavedova:2013} 
\textbf {Della Vedova et al.} ``Geothermal structure of Tyrrhenian Sea. Mar. Geol. 55, 271–289'', 1994
\bibitem{Arisi Rota:2013} 
\textbf {Arisi Rota et Fichera} ``Magnetic interpretation connected to Geomagnetic provinces: the Italian case history. 47th Meeting European Association of Exploration Geophysicists Proc.'', 1985
\bibitem{Buonasorte:2013} 
\textbf {Buonasorte et al.} ``Tectonic structures and geometric setting of the Vulsini Volcanic Complex. Per. Mineral., 56: 123-136'', 1987a
\bibitem{FBarberi:2013} 
\textbf {F. Barberi, Buonasorte G., R. Cioni, Fiordelisi A., Foresi L., S. Iaccarino, Laurenzi MA , Sbrana A., L. Vernia, Villa IM} ``Plio-Plesitocene geological evolution of the geothermal area of Tuscany and Latium. Mem. Descr. Carta Geol. D’It. XLIX (1994), pp 77-134''
\bibitem{Iaccarino:2013} 
\textbf {Iaccarino et al.} ``Osservazioni stratigrafiche sul bordo orientale del Bacino di Radicofani. Mem. Descr. Carta Geol. D’It. XLIX'', 1994
\bibitem{UE:2013}
\textbf{European Commission} ``Map of Solar Irradiation'', 2012
\bibitem{INGV:2004}
\textbf{INGV} ``Seismic Hazard Map (MPS) of Tuscany'', 2004
\bibitem{INGV:2004:ref2}
\textbf{INGV} ``S1-INGV Project (http://esse1.mi.ingv.it/ntc.html)'', 2008
\bibitem{Boccaletti:ref1}
\textbf{Boccaletti, M. \& calamita, F. et all}, "Evoluzione dell'Appennino Tosco-Umbro-Marchigiano durante il Neogene", Giornale di Geologia, Sez.3, vol.48/1-2, 1986
\bibitem{Capecchi:ref1} 
\textbf{Capecchi, F. \& Pranzini, G.}, "Studi geologici e idrogeologici della pianura di Pistoia", Boll.Soc.Geol. It. 94, 1975
\bibitem{Destefani:ref1}
\textbf{De Stefani C.}, "I dintorni di Monsummano e Montecatini in Val di Nievole", Boll. Soc. Geol. It. 42-52, 1887
\bibitem{Frazzuoli:ref1}
\textbf{Frazzuoli, M. \& Maestrelli-Manetti O. }, "I nuclei Mesozoici di Monsummano, Montecatini Terme e Marliana (provincia di Pistoia)", 1973
\bibitem{Ghelardoni:ref1}
\textbf{Ghelardoni, R.}, "Spostamento dallo spartiacque dell'Appennino Settentrionale in corrispondenza di catture idrografiche", Atti Soc. Sc. Nat., Mem. Ser. A, 65, 1958
\bibitem{Arrighi:ref1}
\textbf{Arrighi, A. \& Bertogna, A. \& Naef, S.}, "Montalbano: geologia, flora, fauna, storia e arte", Cnsorzio Interprovinciale per il Montalbano, 1993
\bibitem{Horizon2020:ref1} 
\textbf {Horizon 2020 - Work Programme 2018-2020}, European research infrastructures (including e-Infrastructures), 2016.
\end{thebibliography}
\end{document}